\documentclass[useAMS,usenatbib]{mn2e}
\usepackage{amsmath}
\usepackage{mathabx}
\usepackage{dsfont}
\usepackage{graphicx}
\usepackage{subfigure}

\title[Radial growth of stellar discs]{The instantaneous radial growth rate of stellar discs}
\author[Pezzulli et al.]{G. Pezzulli$^{1}$\thanks{E-mail:
gabriele.pezzulli@unibo.it}; F. Fraternali $^{1,2}$; S. Boissier $^{3,4}$ and J.C. Mu\~noz-Mateos $^{5}$
\\
$^{1}$ University of Bologna, Department of Physics and Astronomy, Viale Berti Pichat 6/2, 40127 Bologna, Italy \\
$^{2}$ University of Groningen, Kapteyn Astronomical Institute, Postbus 800, 9700 AV, Groningen, The Netherlands \\
$^{3}$ Aix Marseille Universit\'e, CNRS, LAM, UMR 7326, 13388 Marseille, France \\
$^{4}$ INAF, Osservatorio Astronomico di Bologna, Via Ranzani, 1, 40127 Bologna, Italy \\
$^{5}$ European Southern Observatory, Alonso de Cordova 3107, Vitacura, Casilla 19001, Santiago, Chile}

\begin{document}

\date{Accepted 2015 May 8.
Received 2015 May 2;
in original form 2015 March 2
}

\pagerange{\pageref{firstpage}--\pageref{lastpage}} \pubyear{2015}

\maketitle

\label{firstpage}

\begin{abstract}
We present a new and simple method to measure the instantaneous mass and radial growth rates of the stellar discs of spiral galaxies, based on their star formation rate surface density (SFRD) profiles. Under the hypothesis that discs are exponential with time-varying scalelengths, we derive a universal theoretical profile for the SFRD, with a linear dependence on two parameters: the specific mass growth rate $\nu_\textrm{M} \equiv \dot{M}_\star/M_\star$ and the specific radial growth rate $\nu_\textrm{R} \equiv \dot{R}_\star/R_\star$ of the disc. We test our theory on a sample of 35 nearby spiral galaxies, for which we derive a measurement of $\nu_\textrm{M}$ and $\nu_\textrm{R}$. 32/35 galaxies show the signature of ongoing inside-out growth ($\nu_\textrm{R} > 0$). The typical derived e-folding timescales for mass and radial growth in our sample are $\sim 10 \; \textrm{Gyr}$ and $\sim 30 \; \textrm{Gyr}$, respectively, with some systematic uncertainties.
More massive discs have a larger scatter in $\nu_\textrm{M}$ and $\nu_\textrm{R}$, biased towards a slower growth, both in mass and size. 
We find a linear relation between the two growth rates, indicating that our galaxy discs grow in size at $\sim 0.35$ times the rate at which they grow in mass; this ratio is largely unaffected by systematics. Our results are in very good agreement with theoretical expectations if known scaling relations of disc galaxies are not evolving with time.

\end{abstract}


\section{Introduction}\label{sec::Introduction}
The theory of cosmological tidal torques (\citealt{Peebles}) predicts the mean specific angular momentum of galaxies to be an increasing function of time. If applied to spiral galaxies, in which stars are mostly distributed on a rotating, centrifugally supported, disc, the theory suggests that the outer parts, with higher specific angular momenta, should form later than the inner ones (\citealt{Larson76}, the so-called \emph{inside-out} formation scenario), also implying that spirals should grow in size while they grow in mass. Apart from this quite general prediction provided by cosmology, the details about how stellar discs form and grow in mass and size are not known from first principles and significant observational effort is still required to shed light on the missing links from structure formation to galaxy formation.

An invaluable input for modelers comes from some simple observed properties of the discs of spiral galaxies, that still wait for a comparatively simple theoretical explanation: among them, the exponential radial structure of galaxy discs (\citealt{Freeman}), though sometimes broken at galaxy edges (e.g. \citealt{PohlenTrujillo06}; \citealt{Erwin+08}), and the fact that they obey simple scaling relations, including the Tully-Fisher relation between rotational velocity and mass (\citealt{TF}, see also the `baryonic Tully-Fisher relation', \citealt{McGaugh}; \citealt{Zaritsky+14}), the Fall relation between angular momentum and mass (\citealt{Fall}; \citealt{RF12}) and a more scattered mass-size relation (e.g. \citealt{Shen+03}; \citealt{Courteau+07}), which can also be seen as a corollary of the other two.

Observations of galaxies at different redshifts indicate that stellar discs have an exponential structure since very early epochs (\citealt{Elmegreen+05}; \citealt{Fathi+12}), while it is less clear whether scaling relations are truly universal or they evolve with cosmic time.
For example, direct measurements of the mass-size relation for disc galaxies at various redshifts has led to claims for little or no evolution (e.g. \citealt{Ravindranath+04}; \citealt{Barden+05}; \citealt{Ichikawa+12})
as well as significant or strong evolution (e.g. \citealt{MaoMoWhite98}; \citealt{Buitrago+08}; \citealt{Fathi+12}). The interpretation and comparison of these pioneering studies is made non trivial by inhomogeneities among observations at different redshifts, as well as differences in sample definitions and analysis techniques (see e.g. \citealt{Lange+15}); also, several subtle issues have been shown to significantly bias the results, most notably the selection effect due to cosmological dimming (e.g. \citealt{Simard+99}) and the evolution of M/L ratios due to evolving stellar populations (e.g. \citealt{Trujillo+06}). The related problem of the possible evolution of the Tully-Fisher relation, which also involves kinematic measurements, is even more complex and controversial (e.g \citealt{Vogt+97}; \citealt{MaoMoWhite98}; \citealt{Miller+11}; \citealt{Miller+12} and references therein).

Since these issues are of extreme importance for our understanding of galaxy evolution, more observational effort is desirable, possibly dealing with multiple independent probes, to unveil the growth of the exponential discs of spiral galaxies. In addition to the crucial, but often challenging, comparison of galaxy properties at different redshifts, indirect information can be gained on the size growth of galaxy discs from the study of their properties in the Local Universe. Efforts in this direction can be split in two categories. The first is the exploitation of fossil signals of the past structure of the disc: most notably, chemical enrichment (e.g. \citealt{BP99}, \citealt{Chiappini}; \citealt{MD05}; \citealt{NO06}), and properties of stellar populations, including colour gradients (e.g. \citealt{BelldeJong00}; \citealt{MacArthur+04}; \citealt{Wang+11}), spectrophotometry (e.g. \citealt{MMIII}; \citealt{CALIFA}) and colour-magnitude diagrams (e.g. \citealt{Williams+09}; \citealt{Gogarten+10}; \citealt{Barker+11}). The second possibility is to look for the \emph{instantaneous} signal of the growth process while it is in act. Spiral galaxies are not just passively evolving stellar systems, but keep forming stars at a sustained rate throughout their evolution (e.g. \citealt{AB09}; \citealt{FT12}; \citealt{Tojeiro+13}). Therefore, the radial distribution of newly born stars is a crucial ingredient for the structural evolution of a stellar disc and it can be used as a clean and direct probe of its growth.

Thanks to the deep UV photometry of the Galaxy Evolution Explorer (GALEX, \citealt{GALEX}), radial profiles of the star formation rate surface density (SFRD) of nearby galaxies can now be traced out to considerable galactocentric distances and low levels of star formation activity. SFRD profiles, as traced by the UV light emitted by young stars, often turn out to be quite regular and, in many cases, remarkably similar to exponentials (\citealt{Boissier+07}, see also \citealt{Goddard+10}). This supports the idea that star formation is tightly linked with whatever process is responsible for the mantainance and evolution of the exponential structure of galaxy discs. A closer inspection of the aforementioned SFRD profiles reveals, in many galaxies, some deviations from the exponential shape, in the form of a central downbending or depletion (see also \citealt{MMIII}). This is also clearly seen in the SFRD of the Milky Way, as traced, for example, by the distribution of Galactic supernova remnants (\citealt{CB98}). Indeed, star formation becoming progressively less effective towards the centre of galaxies is not very surprising, within the inside-out formation scenario: in the inner regions, the bulk of gas accretion and conversion into stars occurs quite early and relatively little residual star formation is expected to be in place there at late epochs, while the outskirts are still in their youth. Ultimately, the observed properties of SFRD profiles of spiral galaxies are in qualitative agreement with the inside-out  paradigm and therefore they are good candidates to enclose the signal of radial growth. The aim of this work is to give a simple quantitative description of this signal and a method to derive a measurement of the instantaneous mass and radial growth rates of the discs of spiral galaxies from the amplitude and shape of their SFRD profiles.

An earlier attempt in this direction has been done by \cite{MM+07}. They assumed that surface densities of both stellar mass and star formation rate can be approximated with exponential profiles, though with different scalelengths, implying exponential profiles for the specific star formation rate (sSFR) as well, with sSFR increasing with radius for inside-out growing galaxies. This parametrization was applied to a sample of nearby spiral galaxies and the results were compared with the predictions of simple structural evolution models, providing constraints on the inside-out process.

In this work we make a step forward, proposing a method that is both simpler and more powerful. Rather than modeling both stellar mass and SFRD with exponentials, we assume that just stellar discs are exponential at any time, with time-varying scalelengths. This naturally brings us (Sec. \ref{sec::theory}) to predict a universal shape for the SFRD profile with the observed properties outlined above, namely an inner depletion and an outer exponential decline. Furthermore, our theoretical profile has a very simple (linear) dependence on the disc mass and radial growth rates; hence, these parameters can be directly derived from observations in a model-independent way. We apply our method to a sample of nearby spiral galaxies described in Sec. \ref{sec::sample}, discuss our analysis in Sec. \ref{sec::analysis} and present our results in Sec. \ref{sec::results}. In Sec. \ref{sec::scalingrelations} the consequences of our findings are discussed on the issue of whether known scaling relations for galaxy discs are evolving with time or not. In Sec. \ref{sec::Summary} we give a summary.

\section{Star formation in exponential discs}\label{sec::theory}
\subsection{A simple model}\label{sec::simplemodel}
Let us assume that the mass surface density $\Sigma_\star$ of the stellar disc of a spiral galaxy is well described, at any time, by an exponentially declining function of radius $R$, identified by a radial scalelength $R_\star$ and a mass $M_\star$ \footnote{Throughout the paper, when referring to a spiral galaxy, we will use the symbol $M_\star$ to denote the stellar mass of its disc component alone.}, both allowed to vary with time $t$:
\begin{equation}\label{expdisc}
\Sigma_\star(t, R) = \frac{M_\star(t)}{2 \pi R_\star^2(t)} \exp \left( -\frac{R}{R_\star(t)} \right)
\end{equation}
Just taking the partial time derivative of \eqref{expdisc} we get a very simple prediction for the star formation rate surface density $\dot{\Sigma}_\star$ as a function of time and galactocentric radius:
\begin{equation}
\dot{\Sigma}_\star(t, R) =  \left( \nu_\textrm{M}(t) + \nu_\textrm{R}(t) \left( \frac{R}{R_\star(t)} - 2 \right) \right) \Sigma_\star(t, R)
\end{equation}
where $\Sigma_\star$ is given by \eqref{expdisc}, while the quantities $\nu_\textrm{M}$ and $\nu_\textrm{R}$ are defined by:
\begin{equation}\label{defnuM}
\nu_\textrm{M}(t) := \frac{d}{dt} \left( \ln M_\star(t) \right) = \frac{\dot{M}_\star(t)}{M_\star(t)}
\end{equation}
\begin{equation}\label{defnuR}
\nu_\textrm{R}(t) := \frac{d}{dt} \left( \ln R_\star(t) \right) = \frac{\dot{R}_\star(t)}{R_\star(t)}
\end{equation}
We discuss them more thoroughly in Sec. \ref{sec::TheoryGrowthRates}.

\subsection{Theoretical \emph{caveats}}\label{sec::TeoCaveats}
At least two \emph{caveats} should be kept in mind when considering the elementary inference outlined in Sec. \ref{sec::simplemodel}. 

First, by identifying $\dot{\Sigma}_\star$ with $(\partial \Sigma_\star / \partial t)$, we have implicitly neglected any contribution coming from a possible net radial flux of stars, that is $1/R \; \partial / \partial R (2 \pi R \Sigma_\star u^R_\star)$, $u^R_\star$ being the net radial velocity of stars. While radial migration of stars is widely recognized to be a fundamental ingredient of galaxy evolution, it has also been shown (\citealt{SB02}; \citealt{Roskar+12}) that its main working mechanism is basically a switch in the radial position of two stars in different circular orbits (the so-called \emph{churning}, \citealt{SB09}). This process produces no dynamical heating, no \emph{net} radial flow of stars and no change in the mass distribution of the disc. Of course, some minor contribution to radial migration are also expected from other processes: breaks to our approximation can be expected in some cases, mostly in the inner regions, where dynamical processes might be associated with the formation of bars, rings and pseudobulges (e.g. \citealt{SellwoodReview}) and at the outer edge, where radial migration has been proposed to induce changes in the outer structure of discs (\citealt*{Yoachim+12}). More complex effects are also possible due to the interplay between stellar dynamics, gas dynamics and star formation; quite different approaches to this problem can be found, for example, in \cite{SB09}, \cite{KPA13} and \cite{MCM14}.

Second, since stellar populations, during their evolution, return a substantial fraction of their mass to the ISM (\citealt{Tinsley80}), it is not necessarily trivial to connect the time derivative $\dot{\Sigma}_\star$ to observed values of SFRD. In the following, we will adopt the instantaneous recycling approximation (IRA) and assume that a constant \emph{return fraction} $\mathcal{R}$ of the mass of a stellar population is instantaneously given back to the ISM. Accordingly, our $\dot{\Sigma}_\star$ represents the \emph{reduced} star formation rate surface density and it is equal to the observed SFRD multiplied by a factor $(1-\mathcal{R})$, although we will often omit the attribute \emph{reduced}, for brevity. More detailed studies (e.g. \citealt{LK11}) show that the majority of the returned mass is released within $\sim 1 \; \textrm{Gyr}$ from the birth of a population; hence, our approximation will be valid in those galaxies, or galaxy regions, where star formation has not been varying abruptly on timescales shorter than $\sim 1 \; \textrm{Gyr}$. 
It can be easily seen that such abrupt changes can, in principle, be taken into account by replacing $\mathcal{R}$ with an \emph{effective return fraction} $\mathcal{R}_\textrm{eff}$, which is higher or lower than $\mathcal{R}$ for abrupt quenching or starbursting events, respectively. These may be due, for instance, to tidal or ram pressure stripping or, viceversa, to significant recent accretion events. Similar effects can sometimes be observationally inferred in low surface brightness regions or in low surface brightness galaxies as a whole (\citealt{Boissier+08}) and may also be related to the phenomenology of extended UV (XUV) discs (\citealt{Thilker+07a}). Unfortunately, in general, neither the magnitude nor the direction of the needed correction can be known \emph{a priori}. However, these possibilities should be kept in mind when considering peculiar features in the observed SFRD profiles of some individual objects.

\subsection{Mass and radial growth rates}\label{sec::TheoryGrowthRates}
The quantities $\nu_\textrm{M}$ and $\nu_\textrm{R}$, defined in \eqref{defnuM} and \eqref{defnuR}, shall be called \emph{specific mass growth rate} and \emph{specific radial growth rate}. The word \emph{specific} refers to the fact that they represent the mass and radial growth rates $\dot{M}_\star$ and $\dot{R}_\star$, \emph{normalized} to the actual value of mass $M_\star$ and scalelength $R_\star$, respectively. However, since in this work we will deal only with specific quantities, we will often omit the attribute and refer to them just as mass and radial growth rates, for brevity. 

While $\nu_\textrm{M}$ is always positive (stellar mass is never destroyed), $\nu_\textrm{R}$ can in principle take both signs, positive values being expected in the case of inside-out growth. At any time, the inverse of $\nu_\textrm{M}$ and $\nu_\textrm{R}$ can be interpreted as instantaneous estimates of the timescales for the growth of the stellar mass and scalelength, respectively (or for disc shrinking, in the case $\nu_\textrm{R} < 0$).

We notice that $\nu_\textrm{M}$ is strictly related to another physical quantity, namely the specific star formation rate (sSFR). More precisely, following the terminology of \cite{Lilly}, $\nu_\textrm{M}$ coincides with the \emph{reduced specific star formation rate} of the disc, where the word \emph{reduced} (which we will omit from now on) refers to the fact that we are including the effect of the return fraction $\mathcal{R}$. Since our $\nu_\textrm{M}$ refers to the disc alone, it should not be confused with the sSFR of a whole spiral galaxy, which is evaluated including also the other stellar components, like the bulge. While the bulge can give a non-negligible contribution to the total stellar mass, it usually harbours little or no star formation: hence, the sSFR of a whole galaxy and of its disc alone can differ significantly (\citealt{Abramson+14}). Also, $\nu_\textrm{M}$ should not be confused with the \emph{local} sSFR $(\dot{\Sigma}_\star/\Sigma_\star)$, which is, in general, a function of galactocentric radius (e.g. \citealt{MM+07}).

The analogous quantity for the stellar scalelength, the (specific) radial growth rate $\nu_\textrm{R}$, has been studied much less (and, as far as we know, not even clearly defined until now). To provide a simple method for its measurement is the main aim of this work. Since $\nu_\textrm{M}$ and $\nu_\textrm{R}$ have the same physical dimensions and refer to the two basic properties of an exponential disc, we are also interested to measure both quantities at the same time and to attempt a comparison between them. This is indeed a natural outcome of our method (Sec. \ref{sec::predictions}) and will bring us to the most interesting consequences of our results (Sec. \ref{sec::scalingrelations}).

\subsection{Predicted properties of SFRD profiles}\label{sec::predictions}
Our simple model predicts that, if a galaxy is observed at some particular time, its SFRD should follow a radial profile of the form:
\begin{equation}\label{teoSFRD}
\dot{\Sigma}_\star(R) =  \frac{M_\star}{2 \pi R_\star^2} \left( \nu_\textrm{M} + \nu_\textrm{R} \left( \frac{R}{R_\star} - 2 \right) \right) \exp \left( -\frac{R}{R_\star} \right)
\end{equation}
\begin{figure}
\centering
\includegraphics[width=8cm]{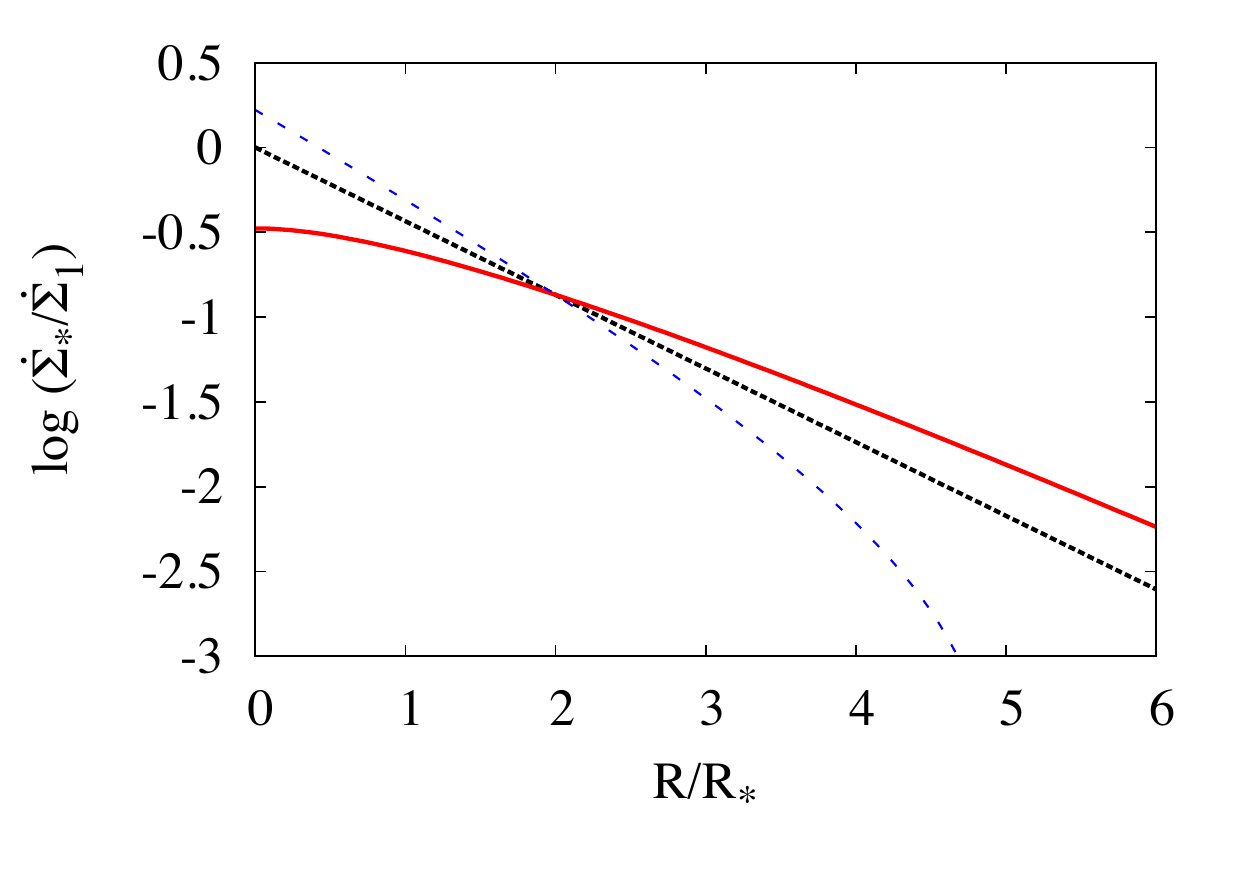}
\caption{Predicted shape of the SFRD profile, in dimensionless units, for some illustrative values of the radial growth rate $\nu_\textrm{R}$: absence of radial growth ($\nu_\textrm{R} = 0$, dotted black line), inside-out growth ($\nu_\textrm{R} = 1/3\;\nu_\textrm{M}$, solid red line) and disc shrinking ($\nu_\textrm{R} = -1/3\;\nu_\textrm{M}$, dashed blue line). The radius is in units of $R_\star$, the SFRD is normalized to $\dot{\Sigma}_1 \equiv \nu_\textrm{M}M_\star/2 \pi R_\star^2$, so that the comparison refers to discs with the same stellar mass, scalelength and \emph{global} sSFR.}\label{fig::teoSFRD}
\end{figure}
In Fig. \ref{fig::teoSFRD} the predicted shape of the SFRD profile is drawn out, in dimensionless units, for some representative situations, which differ for the sign of the radial growth parameter $\nu_\textrm{R}$. To achieve a fully dimensionless description of the model, we use here, as a parameter, the dimensionless ratio $\nu_\textrm{R}/\nu_\textrm{M}$, which has the same sign of $\nu_\textrm{R}$, since, as pointed out in Sec. \ref{sec::TheoryGrowthRates}, $\nu_\textrm{M}$ is always positive. Also, with our adopted dimensionless units, the curves in Fig. \ref{fig::teoSFRD} compare with each other as model discs that share the same mass,  scalelength and \emph{global} sSFR $\nu_\textrm{M}$, but differ in the spatial distribution of star formation, depending on the presence (and the direction) of an evolution of the scalelength with time. 

In the absence of radial evolution ($\nu_\textrm{R} = 0$), the scalelength of the stellar disc is constant with time, stars always form with the same spatial distribution and the SFRD profile is an exponential as well, with the same scalelength of the already present stellar disc. In the case of disc shrinking ($\nu_\textrm{R} < 0$) star formation is enhanced, with respect to the previous case, in the inner regions, but it abruptly drops in the outskirts (reaching a vertical asymptote at $R/R_\star = (2 + \nu_\textrm{M}/|\nu_\textrm{R}|)$). Conversely, for inside-out growth ($\nu_\textrm{R} > 0$), the SFRD shows the characteristic depletion in the central regions, while it is enhanced in the outskirts, where it ultimately gently declines with increasing radius, with an asymptotic behaviour, at large radii, similar to the one of the stellar mass distribution.

\section{Sample and data}\label{sec::sample}
\subsection{Sample definition}
To define our sample, we started from the one studied by \cite{MMIII}, which consists of 42 nearby spiral galaxies observed both by \emph{Spitzer} and by GALEX, in the context of the SINGS survey (\citealt{SINGS}). For these and other nearby galaxies, radial profiles have been derived and published by \cite{MMI} and \cite{MMII}, for multi-wavelength emission ranging, for most galaxies, from FIR to FUV. Such a broad range is useful to trace both the stellar mass and the star formation rate, corrected for the effect of dust extinction. \cite{MMII} give radial profiles for the extinction in the UV, as inferred from TIR/UV flux ratios.

For this work, we made use of the radial profiles of the emission in the FUV GALEX band, corrected for extinction in the FUV ($A_\textrm{FUV}$). We also used profiles of emission in the $3.6 \; \mu m$ IRAC band, which we assume to be a good tracer of the stellar mass surface density. Some contamination may arise from the $3.3 \; \mu\textrm{m}$ PAH line, hot dust and AGB stars; however, these contributions are only expected to be important at the small scales of individual star-forming regions and just a mild effect persists at larger scales (\citealt{Meidt+12}). Furthermore, dust extinction at this wavelength is negligible and the M/L is quite insensitive to variations in age and metallicity, if compared to the optical bands (\citealt{Meidt+14}).

From the original sample of 42 galaxies, we excluded 5 galaxies (NGC 3049, NGC 3938, NGC 4254, NGC 4321, NGC 4450) for which FUV meaurements were lacking because the GALEX FUV detector was turned off for technical reasons, one galaxy (NGC 7552) which was not present in the \cite{MMII} sample and another one (NGC 4625) for which the $A_\textrm{FUV}$ radial profile was constituted of only one point. Therefore we ended up with a final sample of 35 galaxies.
We have considered some possible additional criteria to further restrict our sample. Since our main goal is the study of the slow, continuous, evolution of the regular exponential structure of discs, galaxies that are suspected to be undergoing violent transient events, like interactions or mergers, could be excluded from the analysis. Furthermore, since we make a quantitative analysis of azimuthally averaged radial profiles, we could exclude those galaxies for which the geometrical parameters involved in the average (inclination and position angle) are not known with good accuracy or are suspected to be varying with radius.
At least 3 galaxies (NGC 1097, NGC 1512, NGC 5194) have a nearby companion and for at least one (NGC 1512) the adopted position angle reproduces the outer isophotes better than the inner ones. However, it is not clear whether such selections could be done in a completely unbiased way. Also considering that our sample is relatively small, we decided to homogeneously analyse the whole set of 35 galaxies. Nonetheless, the aforementioned \emph{caveats} should be kept in mind while considering our results. We adopt morphological classifications, distances and inclinations (as derived from axis ratios) from \cite{MMIII}; these properties can also be found here in Table \ref{table::sample}.

\begin{table*}
 \centering
  \caption{Basic properties of galaxies in our sample. Morphological classification, distances and inclinations (derived from axis ratios) are as in Mu\~noz-Mateos et al. (2011).} \label{table::sample}
  \begin{tabular}{lcccccc}
  \hline
   Galaxy & RA (J2000) & Dec (J2000) & Morphological & Type & Distance & $\cos i$\\
& $h$ $m$ $s$ & $^{\degree}$ $^{\prime}$ $^{\prime \prime}$ & Type & T & (Mpc) & \\
 \hline
NGC 0024 & 00 09 56.5 & -24 57 47.3 & SA(s)c & 5 & 8.2 & 0.224\\
NGC 0337 & 00 59 50.1 & -07 34 40.7 & SB(s)d & 7 & 25 & 0.621 \\
NGC 0628 & 01 36 41.8 & 15 47 00.5 & SA(s)c & 5 & 11 & 0.905 \\
NGC 0925 & 02 27 16.9 & 33 34 45.0 & SAB(s)d & 7 & 9.3 & 0.562 \\
NGC 1097 & 02 46 19.1 & -30 16 29.7 & SB(s)b & 3 & 15 & 0.677 \\
NGC 1512 & 04 03 54.3 & -43 20 55.9 & SB(r)a & 1 & 10 & 0.629 \\
NGC 1566  & 04 20 00.4 & -54 56 16.1 & SAB(s)bc & 4 & 17 & 0.795 \\
NGC 2403 & 07 36 51.4 & 65 36 09.2 & SAB(s)cd & 6 & 3.2 & 0.562 \\
NGC 2841  & 09 22 02.6 & 50 58 35.5 & SA(r)b & 3 & 14 & 0.432 \\
NGC 2976 & 09 47 15.5 & 67 54 59.0 & SAc pec & 5 & 3.6 & 0.458 \\
NGC 3031 & 09 55 33.2 & 69 03 55.1 & SA(s)ab & 2 & 3.6 & 0.524 \\
NGC 3184  & 10 18 17.0 & 41 25 28.0 & SAB(rs)cd & 6 & 8.6 & 0.932 \\
NGC 3198 & 10 19 54.9 & 45 32 59.0 & SB(rs)c & 5 & 17 & 0.388 \\
IC 2574 & 10 28 23.5 & 68 24 43.7 & SAB(s)m & 9 & 4.0 & 0.409 \\
NGC 3351 & 10 43 57.7 & 11 42 13.0 & SB(r)b & 3 & 12 & 0.676 \\
NGC 3521  & 11 05 48.6 & -00 02 09.1 & SAB(rs)bc & 4 & 9.0 & 0.464 \\
NGC 3621 & 11 18 16.5 & -32 48 50.6 & SA(s)d & 7 & 8.3 & 0.577 \\
NGC 3627 & 11 20 15.0 & 12 59 29.6 & SAB(s)b & 3 & 9.1 & 0.462 \\
NGC 4236 & 12 16 42.1 & 69 27 45.3 & SB(s)dm & 8 &4.5 & 0.329 \\
NGC 4536 & 12 34 27.1 & 02 11 16.4 & SAB(rs)bc & 4 & 15 & 0.421\\
NGC 4559 & 12 35 57.7 & 27 57 35.1 & SAB(rs)cd & 6 & 17 & 0.411 \\
NGC 4569 & 12 36 49.8 & 13 09 46.3 & SAB(rs)ab & 2 & 17 & 0.463 \\
NGC 4579 & 12 37 43.6 & 11 49 05.1 & SAB(rs)b & 3 & 17 & 0.797 \\
NGC 4725 &12 50 26.6 & 25 30 02.7 & SAB(r)ab pec & 2 & 17 & 0.710 \\
NGC 4736 & 12 50 53.1 & 41 07 13.6 & (R)SA(r)ab & 2 & 5.2 & 0.813 \\
NGC 4826 & 12 56 43.8  & 21 40 51.9 & (R)SA(rs)ab & 2 & 7.5 & 0.540 \\
NGC 5033 & 13 13 27.5 & 36 35 38.0 & SA(s)c & 5 & 13 & 0.467 \\
NGC 5055 & 13 15 49.3 & 42 01 45.4 & SA(rs)bc & 4 & 8.2 & 0.571 \\
NGC 5194 & 13 29 52.7 & 47 11 42.6 & SA(s)bc pec & 4 & 8.4 & 0.804 \\
NGC 5398$^a$
 & 14 01 21.6 & -33 03 49.6 & (R')SB(s)dm pec & 8.1 & 16 & 0.607 \\
NGC 5713 & 14 40 11.5 & -00 17 21.2 & SAB(rs)bc pec & 4 & 27 & 0.893 \\
IC 4710 & 18 28 38.0 & -66 58 56.0 & SB(s)m & 9 & 8.5 & 0.778 \\
NGC 6946 & 20 34 52.3 & 60 09 14.2 & SAB(rs)cd & 6 & 5.5 & 0.852 \\
NGC 7331 & 22 37 04.1 & 34 24 56.3 & SA(s)b & 3 & 15 & 0.352 \\
NGC 7793 & 23 57 49.8 & -32 35 27.7 & SA(s)d & 7 & 3.9 & 0.677 \\
\hline
\end{tabular} \\
$(a)$ In the original sample, this galaxy was referred to as TOL 89, which is the name of an HII region embedded within it.
\end{table*}

\subsection{Stellar mass surface density}\label{sec::DataStars}
To get the stellar mass distribution, we made use of the high resolution (6 arcsec) radial profiles at 3.6 $\mu$m from \cite{MMI}. To convert from surface brightness to mass surface density, we used the conversion formula:
\begin{equation}
\frac{\Sigma_\star}{\textrm{M}_\Sun \; \textrm{pc}^{-2}} = 1.9 \cdot 10^7 \; \cos i \; \frac{I_{3.6 \mu\textrm{m}}}{\textrm{Jy} \; \textrm{arcsec}^{-2}}
\end{equation}
which is the one derived by \cite{Leroy+08}, though written in different units and modified for a K-band mass-to-light ratio equal to 0.8, instead of 0.5 \footnote{The formula of \cite{Leroy+08} was based on their measured linear $3.6 \; \mu\textrm{m}$-to-K-band flux conversion and on an assumed K-band mass-to-light ratio. Changing the latter from 0.5 to 0.8 is equivalent to introducing an additional factor 1.6 in the $3.6 \; \mu\textrm{m}$-to-stellar mass conversion.\label{footnote::ML}}. This ratio is subject to several uncertainties (e.g. \citealt{Bell+03}); our choice was made to maximize consistency with the previous work by \cite{MM+07}. We discuss the consequences of this and other systematics in Sec. \ref{sec::systematics}.

\subsection{Star formation rate surface density}\label{sec::DataSFRD}
To derive the SFRD profiles, we took the low resolution (48 arcsec) radial profiles in the FUV band from \cite{MMII} and corrected them for extinction using the $A_\textrm{FUV}$ radial profiles at the same resolution. In that work, two estimates of $A_\textrm{FUV}$ are provided, based on two slightly different dust attenuation prescriptions by \cite{B05} and by \cite{C08}, the latter containing a refinement to take additional dust heating from old stars into account. In this work we used extinction profiles from the \cite{B05} prescription. We made this choice to maximize simplicity and reproducibility of our analysis (this recipe does not require additional information on K-band photometry). \cite{MMII} showed that the two prescriptions differ significantly only for early-type galaxies (ellipticals and lenticulars), which are absent in our sample.We verified that our conclusions are not modified when changing the adopted prescription.

The extinction corrected profile $\mu_\textrm{FUV,corr}$ (expressed in the AB magnitude system) was converted into a SFRD by making use of the formula:
\begin{equation}\label{SFRDconv}
\frac{\dot{\Sigma}_\star}{\textrm{M}_\Sun \; \textrm{pc}^{-2} \; \textrm{Gyr}^{-1}} = (1-\mathcal{R})\cos i \; 10^{ -0.4 \mu_\textrm{FUV,corr} + 10.413}
\end{equation}
which is again consistent with \cite{MM+07}. For the return fraction, we adopted $\mathcal{R} = 0.3$, which is intermediate between possible values for different IMF choices (\citealt{LK11}; \citealt{FT12}). Also the systematics associated with \eqref{SFRDconv} is discussed in Sec. \ref{sec::systematics}.

\begin{figure*}
\centering
\subfigure{\includegraphics[width=16cm]{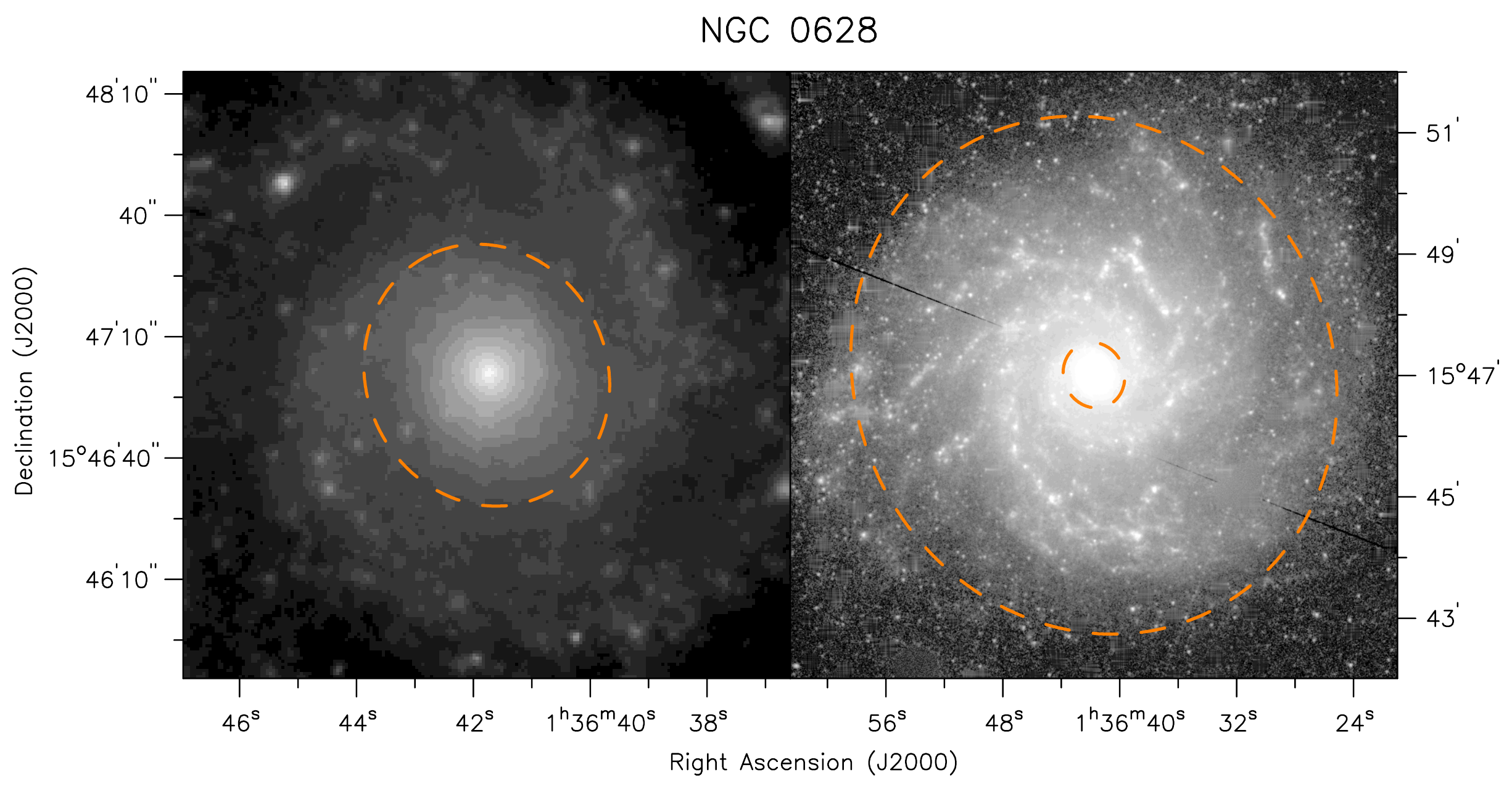}}
\subfigure{\includegraphics[height=6cm]{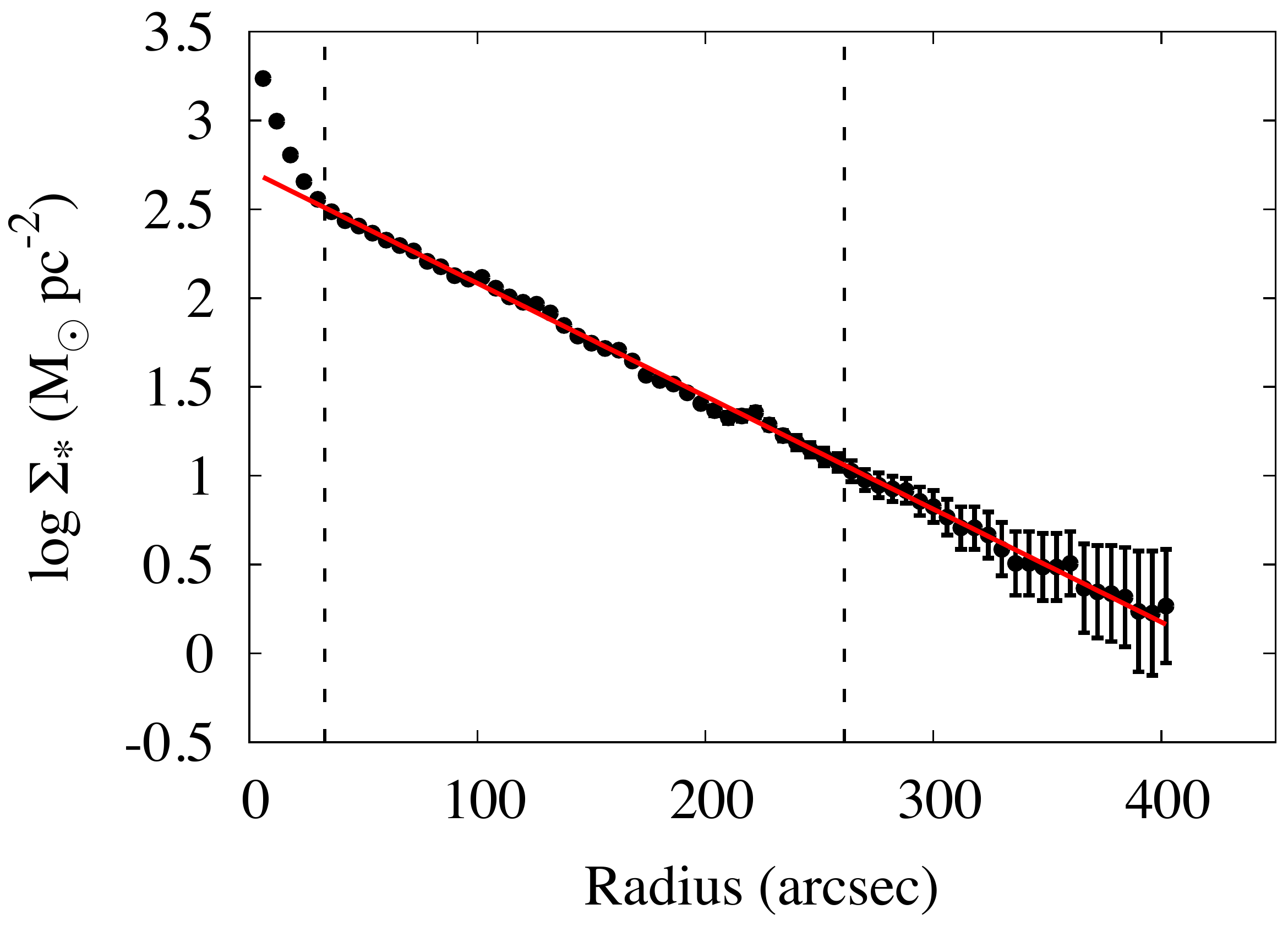}}
\subfigure{\includegraphics[height=6cm]{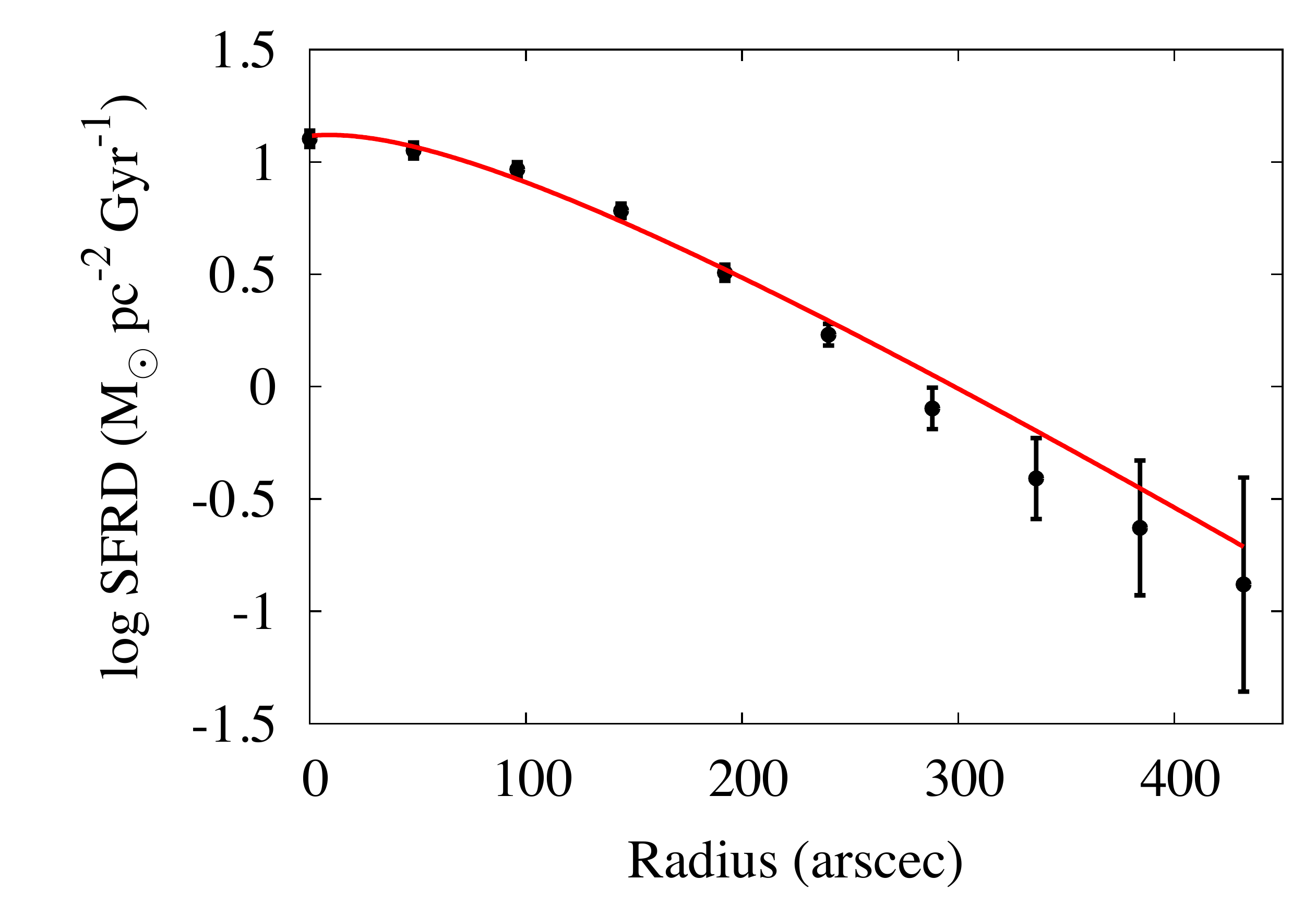}}
\caption{Our analysis for the galaxy NGC 628. \emph{Upper panels} Selection of the domain where the emission at $3.6 \; \mu\textrm{m}$ is dominated by the light from the stars in the the disc: \emph{left} the inner ellipse at 33 arcsec, out of which the spiral structure appears, \emph{right} the outer ellipse at 261 arcsec, out of which the contribution from noise becomes significant. Note that the two images have very different scale and contrast. The inner ellipse is also shown in the right panel, to make the whole selected region visible at once. \emph{Lower-left panel} Exponential fit to the stellar mass surface density, as traced by the emission at $3.6 \; \mu\textrm{m}$. The vertical dashed lines mark the limits of the domain that we have selected for this fit; in this case, the best-fitting exponential also extends further out in the outskirts. \emph{Lower-right panel} Fit of our theoretical SFRD profile to the one obtained from extinction-corrected FUV light; a visual comparison with Fig. \ref{fig::teoSFRD} is already enough to recognize this as an inside-out growing galaxy.}\label{fig::NGC0628}
\end{figure*}
\section{Analysis}\label{sec::analysis}
For each galaxy in the sample, we performed our analysis in two steps. First, we made an exponential fit to the radial profile of the stellar mass surface density of the disc (see Sec. \ref{sec::DataStars}), deriving the values for the disc mass $M_\star$ and scalelength $R_\star$. Then, keeping these parameters fixed, we fitted our theoretical profile \eqref{teoSFRD} to the SFRD data (see Sec. \ref{sec::DataSFRD}). This second fit is the test bed for our theory. If successful, it provides our measurement of the two disc growth parameters: the specific mass growth rate $\nu_\textrm{M}$ and the specific radial growth rate $\nu_\textrm{R}$.

A more extended description of the two steps is given in Secc. \ref{sec::FitStars} and \ref{sec::FitSFRD}. They are depicted, for each individual galaxy, in an Atlas, which we provide as supplementary online material. A representative example, for the galaxy NGC 628, is reported here for illustrative purposes (Fig. \ref{fig::NGC0628}).

All our fits were performed with a standard Marquardt-Levenberg algorithm. We also repeated the whole analysis with a different method (see Sec. \ref{sec::fittingstrategy}) and verified that our results are robust with respect to the fitting strategy.

\subsection{Fit of exponential discs}\label{sec::FitStars}
In order to extract the disc parameters from the radial profiles described in Sec. \ref{sec::DataStars}, we performed a simple exponential fit, for each galaxy, on a radial domain where the NIR emission is dominated by the disc component.

Such a domain was identified, on a case-by-case basis, considering the shape of the $3.6 \; \mu \textrm{m}$ profile with the aid of the direct visual inspection of the 2D maps at the same wavelength. For details about how these maps were obtained the reader is referred to \cite{Regan+04}, \cite{Dale+05} and \cite{MMI}. For each galaxy, the minimum and the maximum radius of our selected domain are the semi-major axes of two concentric ellipses, with centre and orientation equal to the ones used in the derivation of the profiles. The inner ellipse was chosen to exclude the central bright component, if present, like a bulge, a bar, or a central ring; the detection of spiral arms has been used in some cases as an evidence for the prominence of the disc component in a given region. In 4 cases (NGC 2403, IC 2574, NGC 4236, NGC 4826) we have found that the adopted centre of the ellipses did not coincide with the peak of the 3.6 $\mu$m emission; for these galaxies, an inner ellipse was selected with a semi-major axis equal or greater than twice the observed offset. The outer ellipse was most of the times selected to exclude those external regions where a contribution to the emission coming from the noise was found to be significant; for our data, this happens at a typical value of $\log(I_{3.6\;\mu\textrm{m}}/\textrm{Jy}\;\textrm{arcsec}^{-2}) \sim -6.5$.

In 5 galaxies (NGC 3521, NGC 3621, NGC 4736, NGC 5055, NGC 7331) we found a significant flattening of the 3.6 $\mu$m profile well above the noise level and we excluded the outer region from the exponential fit for these objects. The most striking case is NGC 4736, where the effect is probably related to the presence of a prominent outer ring. In 3 of the above cases (NGC 3521, NGC 3621, NGC 5055) the change of slope occurs very near to the outermost radius where a regular spiral pattern can be seen. In the remaining two objects (NGC 3521 and NGC 7331) the flattening is associated with an abrupt change in the geometry of the $3.6\;\mu\textrm{m}$ emission, with the isophotes becoming remarkably large and irregular in the outer regions. We ignore the physical origin of this effect; nonetheless, these peculiarities should be kept in mind in the interpretation of our results for these objects. 

In most cases, our fits were performed weighting each point according to the nominal error, quoted in the original profiles. However, this is not necessarily always the best choice. Real galaxies are not expected to precisely follow an exponential, since transient perturbations like spiral arms, which are by definition ubiquitous in spiral galaxies, can sometimes overimpose oscillations on an underlying regular disc. This effect can become particularly important in the presence of spiral arms with a small pitch angle, tending to dominate the emission in a limited radial range. If, for a given galaxy, points with small error bars happen to be preferentially located in a region dominated by spiral structure, the formal best-fitting profile will be biased to reproduce transient features, potentially missing the overall structure of the disc. For 7 galaxies (NGC 1097, NGC 1512, NGC 3031, NGC 3184, NGC 3351, NGC 4569, NGC 7793) we have found that an unweighted fit provided a better description of the overall structure of the profiles in the considered radial range.

Although the whole procedure is slightly subjective, we verified that it gave a better account to the observed properties of our galaxies, in the domain of interest, with respect to a more complex global analysis, involving more components and parameters. This approach is the most suitable for our purposes, since we are just interested to reliably derive the disc parameters, rather than to get a detailed structural decomposition of the whole galaxy. We verified that the parameters we found were stable with respect to small variations in the selection of the inner and outer radii.

An example of our domain selection and disc-fitting procedure is given in Fig. \ref{fig::NGC0628} (upper panels and lower-left panel) for the case of NGC 628; similar images and plots for the other galaxies can be found in the online Atlas.

\subsection{Fit of the star formation rate surface density}\label{sec::FitSFRD}
In the second step of our analysis, we fitted equation \eqref{teoSFRD} to the observed SFRD profiles, keeping fixed the structural parameters $M_\star$ and $R_\star$ found in the previous step.

The SFRD profiles have a worst spatial resolution, and hence a more limited number of independent points, with respect to the mass surface density profiles. As a consequence, the results of the SFRD fits are more sensitive to changes in the adopted radial domain. In order to limit the dependence of our analysis on subjective choices, we decided to always perform the fit on the whole available domain. Not to exclude any point from the inner regions is equivalent to assume that the bulk of star formation is everywhere associated with the disc component. In other words, we neglected possible star formation activity directly occurring in the bulges, which is quite reasonable since these structures are known to be dominated by old stellar populations. Neither we put outer limits to our domain, implying that we did not try to model possible transient star formation episodes that might dominate the UV emission in the outer regions, nor any kind of structural irregularity and, most noticeably, the possible presence of warps, which are generally expected in the periphery of discs (\citealt{Briggs90}). All the SFRD fits were performed weighting points with their nominal errors, which were derived just propagating the errors in the $\mu_\textrm{FUV}$ and $A_\textrm{FUV}$ profiles.

In Fig. \ref{fig::NGC0628} (lower-right panel) the best-fitting SFRD profile is reported for the galaxy NGC 628; similar plots are reported for all galaxies in the online Atlas.

In considering this part of the analysis, it should be kept in mind that, while the parameters $\nu_\textrm{M}$ and $\nu_\textrm{R}$, in the theoretical SFRD profile \eqref{teoSFRD}, are allowed to change in the fitting process, the global slope is strongly constrained by the parameter $R_\star$, which is held fixed to the value previously obtained from the structural fit (Sec. \ref{sec::FitStars}). Hence, notwithstanding the presence of 2 free parameters, we are by no means able to reproduce arbitrary profiles and the fact that we can recover the majority of SFRD distributions shall be regarded as a success of the model and gives us confidence on the meaningfulness of the resulting best-fit parameters.

In 5 cases (NGC 1512, NGC 3521, NGC 3621, NGC 4736, NGC 5055) we clearly detect an outer flattening in the radial profiles of both SFRD and stellar mass surface density. This can be considered as an indication for the existence of a distinct, relatively long-lived, outer star forming component. More detailed studies would be necessary to clarify this point. However, we notice here that three of these galaxies (NGC 1512, NGC 3621, NGC 5055) have been classified by \cite{Thilker+07b} as having a Type 1 XUV disc. We also notice that sometimes (e.g. for NGC 3621) the spatial coincidence between the two breaks is perfect, while in other cases (NGC 3521 and NGC 5055) the SFRD flattening occurs at larger radii, maybe challenging the idea of a common origin of the two phenomena. A unique case is the one of NGC 7331, which has a very prominent flattening of the stellar mass distribution, but an almost exponential SFRD profile, which our model is unable to account for; we refer the reader to \cite{Thilker+07b} and \cite{Ludwig+12} for more specific studies on this peculiar object and its surroundings. Finally, we report one case (IC 4710) where a quite marked downbending is found, at the same radius, in both profiles. This is not a very secure result, since the break occurs out of our chosen outer ellipse (see Sec. \ref{sec::FitStars}); if confirmed, it may be an example of radial migration in the presence of an outer cut-off in star formation efficiency, as described e.g. by \cite{Yoachim+12} (see also Sec. \ref{sec::TeoCaveats}).

\subsection{A note on the fitting strategy}\label{sec::fittingstrategy}
Our choice of separating the analysis in two steps (Secc. \ref{sec::FitStars} and \ref{sec::FitSFRD}) is motivated by the fact that the structural parameters $(M_\star, R_\star)$ physically describe the mass distribution of stellar discs and hence, in principle, they are best measured on the basis of available data for $\Sigma_\star$ alone, irrespective of the distribution of newly born stars. Also, once such a measurement has been achieved, the fact that the SFRD profiles can be reproduced without a further tuning of $(M_\star, R_\star)$ provides a valuable test for the validity of our theory.

However, we also investigated whether our results would change if our 4 parameters $(M_\star, R_\star, \nu_\textrm{M}, \nu_\textrm{R})$ were allowed to vary simultaneously to reproduce both the stellar mass and the SFRD radial profiles. To this purpose, we ran, for each galaxy, a Monte Carlo Markov Chain based on the combined likelihood of both our datasets (Secc. \ref{sec::DataStars} and \ref{sec::DataSFRD}). We then compared the resulting radial growth rates with the ones derived with our preferred strategy, finding an excellent agreement, with a median absolute difference of just $2\cdot10^{-4}\; \textrm{Gyr}^{-1}$. We found some discrepancy in just 4 cases, 2 of which within 2$\sigma$ (NGC 1097 and NGC 7793) and the other 2 within 3$\sigma$ (NGC 3184 and NGC 3351). Note that all these objects belong to the group for which an unweighted fit was found to provide a better description of the overall disc structure (see Sec. \ref{sec::FitStars}), while the effect of oscillations induced by spiral structure was not taken into account in the MCMC experiment. This probably explains even the moderate discrepancies for this small subset. Furthermore, it shows that our partially subjective choice of the weights, discussed in Sec. \ref{sec::FitStars}, has a very limited impact on our results.

\subsection{Notes on systematics}\label{sec::systematics}
We distinguish between two kinds of systematics, those affecting individual galaxies in a different way and those affecting the whole sample more or less homogenously.

To the first group belong distance and inclination. Distance uncertainties affect the physical values of the derived mass and scalelengths, while the inclination uncertainty mainly affects the determination of the mass. Inclination also affects the normalization of both the stellar mass and SFRD profiles, but it does it exactly in the same way; it is easily seen that this implies a vanishing net effect on the estimates of $\nu_\textrm{M}$ and $\nu_\textrm{R}$, which are also, even more obviously, completely independent on the adopted distance. The important consequence of this is that our method allows us to measure the specific mass and radial growth rates of discs with greater accuracy than the mass and scalelength themselves, a fact that we will further exploit in Sec. \ref{sec::scalingrelations}.

The second group of systematics comprises the mass-to-light ratio, the calibration of the FUV-to-SFRD conversion and the return fraction $\mathcal{R}$.
Apart from second order effects, like possible variations of the mass-to-light ratio and the IMF with radius or morphological type, the main uncertainty coming from these systematics is a common multiplicative factor for both the growth parameters, $\nu_\textrm{M}$ and $\nu_\textrm{R}$, for the whole sample, or, in other words, a possible global rescaling of all the derived timescales.
As examples of global systematics, we consider in some more detail the effect of the IMF and of the return fraction. In our calibrations, we have implicitly adopted a Salpeter IMF. To switch, for instance, to the more popular \cite{Kroupa2001} IMF, we should divide the M/L ratio and hence all stellar mass surface densities by a factor 1.6 (see footnote \ref{footnote::ML}, Sec. \ref{sec::DataStars}), while multiplying all star formation rate surface densities by a factor 0.63 (\citealt{KE12}). The net result on the sSFR (and hence on the estimates of $\nu_\textrm{M}$ and $\nu_\textrm{R}$) is less than 1 \%. This is due to the fact that in both cases the impact of the IMF is essentially driven by a common change in normalization associated to the contribution of very low mass stars. The effect of the return fraction $\mathcal{R}$ is stronger: for instance, changing our adopted $\mathcal{R} = 0.3$ into $\mathcal{R}= 0.48$ (which is the largest of the values suggested by \citealt{LK11}) would imply a reduction of all growth rates by a factor 1.35 and an equal increase of all timescales. Unfortunately, the return fraction is a quite uncertain parameter, since it is significantly affected not only by the IMF, but also by the details of the final-to-initial mass relation, which is very difficult to determine observationally.
However, we stress that the dimensionless ratio between $\nu_\textrm{R}$ and $\nu_\textrm{M}$ is unaffected by any of the systematics we have discussed so far. The importance of this fact will be highlighted in Sec. \ref{sec::scalingrelations}.

\begin{table*}
 \centering
  \caption{Best fit structural ($M_\star$ and $R_\star$) and growth ($\nu_\textrm{M}$ and $\nu_\textrm{R}$) parameters for galaxies in our sample. Formal fitting errors are reported, not including contributions due to distance, inclination and calibrations of conversion fomulae. Compared with stellar mass and scalelength, the growth parameters $\nu_\textrm{M}$ and $\nu_\textrm{R}$ are less affected by systematic effects (see Sec. \ref{sec::systematics}). Out of 35 studied galaxies, 32 have a positive radial growth rate $\nu_\textrm{R}$. }\label{table::results}
  \begin{tabular}{lcccc}
  \hline
   Galaxy & $M_\star$ & $R_\star$ & $\nu_\textrm{M}$ & $\nu
 _\textrm{R}$ \\
 & $(10^{9} \; \textrm{M}_\Sun)$ &  (kpc) & $(10^{-2} \; \textrm{Gyr}^{-1})$ & $(10^{-2} \; \textrm{Gyr}^{-1})$ \\
 \hline
NGC 0024 & $3.01 \pm 0.13$ & $1.62 \pm  0.02$ & $5.85 \pm 1.03$ & $2.51 \pm 0.52$ \\
NGC 0337 & $27.7 \pm 2.3$ & $2.15 \pm 0.06$ & $11.6 \pm 3.5$ & $4.95 \pm 1.72$ \\
NGC 0628 & $43.7 \pm 1.4$ & $3.64 \pm 0.05$ & $8.22 \pm 0.36$ & $2.87 \pm  0.21$ \\
NGC 0925 & $11.3 \pm 0.8$ & $3.97 \pm 0.11$ & $10.3 \pm 0.5$ & $0.799 \pm 0.381$ \\
NGC 1097 & $68.2 \pm 8.1$ & $6.32 \pm 0.23$ & $8.74 \pm 1.04$ & $-2.05 \pm 0.75$ \\
NGC 1512 & $14.7 \pm 2.2$ & $2.22 \pm 0.09$ & $3.80 \pm 0.81$ & $1.22 \pm 0.45$ \\
NGC 1566 & $78.0 \pm 5.3$ & $3.30 \pm 0.07$ & $8.21 \pm 0.83$ & $2.90 \pm 0.44$ \\
NGC 2403 & $7.19 \pm 0.22$ & $1.51 \pm 0.02$ & $9.91 \pm 0.25$ & $2.93 \pm 0.16$ \\
NGC 2841 & $92.8 \pm 3.4$ & $ 3.69 \pm 0.05$ & $1.62 \pm 0.08$ & $0.612 \pm 0.045$ \\
NGC 2976 & $2.25 \pm 0.21$ & $0.802 \pm 0.028$ & $5.88 \pm 0.57$ & $2.04 \pm 0.33$ \\
NGC 3031 & $48.3 \pm 2.5$ & $2.54 \pm  0.03$ & $1.99 \pm 0.20$ & $0.750 \pm 0.118$ \\
NGC 3184 & $17.2 \pm 1.6$ & $2.42 \pm 0.08$ & $6.00 \pm 0.80$ & $1.50 \pm 0.55$ \\
NGC 3198 & $32.0 \pm 1.8$ & $3.65 \pm 0.07$ & $8.05 \pm 0.81$ & $3.30 \pm 0.42$ \\
IC 2574 & $1.21 \pm 0.07$ & $3.01 \pm 0.08$ & $12.0 \pm 1.1$ & $3.71 \pm 0.70$ \\
NGC 3351& $45.3 \pm 3.4$ & $2.86 \pm 0.05$ & $3.62 \pm 0.17$ & $0.384 \pm 0.109$ \\
NGC 3521& $54.2 \pm 2.3$ & $1.85 \pm 0.03$ & $4.75 \pm 0.56$ & $1.95 \pm 0.29$ \\
NGC 3621& $24.8 \pm 1.0$ & $1.74 \pm 0.02$ & $9.84 \pm 2.05$ & $4.18 \pm 1.04$ \\
NGC 3627 & $58.2 \pm 2.5$ & $2.36 \pm 0.03$ & $4.66 \pm 0.36$ & $1.24 \pm 0.21$ \\
NGC 4236 & $1.83 \pm 0.10$ & $2.77 \pm 0.07$ & $11.4 \pm 0.5$ & $4.70 \pm  0.27$ \\
NGC 4536 & $27.0 \pm 1.7$ & $3.90 \pm 0.09$ & $11.2 \pm 0.4$ & $1.17 \pm 0.24$ \\
NGC 4559 & $42.7 \pm 1.4$ & $4.47 \pm 0.05$ & $10.6 \pm 1.0$ & $4.10 \pm 0.51$ \\
NGC 4569 & $78.3 \pm 3.3$ & $4.38 \pm 0.05$ & $2.07 \pm 0.11$ & $-0.02940 \pm 0.0781$ \\
NGC 4579 & $87.6 \pm 2.2$ & $3.56 \pm 0.03$ & $1.70 \pm 0.24$ & $0.357 \pm 0.155$ \\
NGC 4725 & $117 \pm 11$ & $5.45 \pm 0.18$ & $2.17 \pm 0.25$ & $0.831 \pm 0.155$ \\
NGC 4736 & $22.8 \pm 3.1$ & $1.13 \pm 0.05$ & $4.75 \pm 1.06$ & $1.18 \pm 0.52$ \\
NGC 4826 & $51.0 \pm 1.3$ & $1.96 \pm 0.01$ & $1.55 \pm 0.21$ & $-0.0850 \pm 0.1068$ \\
NGC 5033 & $23.5 \pm 2.6$ & $3.91 \pm 0.15$ & $10.2 \pm 0.7$ & $2.09 \pm 0.42$ \\
NGC 5055 & $57.4 \pm 3.8$ & $2.50 \pm 0.05$ & $4.87 \pm 0.63$ & $1.45 \pm 0.34$ \\
NGC 5194 & $77.5 \pm 6.1$ & $2.75 \pm 0.07$ & $7.56 \pm 0.55$ & $1.26 \pm 0.38$ \\
NGC 5398& $4.86 \pm 0.22$ & $1.95 \pm 0.03$ & $8.77 \pm 0.19$ & $3.57 \pm 0.10$ \\
NGC 5713 & $54.8 \pm 6.6$ & $1.94 \pm 0.06$ & $11.0 \pm 3.6$ & $4.54 \pm 1.80$ \\
IC 4710 & $2.71 \pm 0.22$ & $2.13 \pm 0.08$ & $7.61 \pm 1.69$ & $1.97 \pm 1.04$ \\
NGC 6946 & $46.6 \pm 2.2$ & $2.67 \pm 0.05$ & $8.31 \pm 0.54$ & $0.765 \pm 0.529$ \\
NGC 7331 & $123 \pm 13$ & $2.64 \pm 0.08$ & $10.5 \pm 3.8$ & $5.02 \pm 1.88$ \\
NGC 7793 & $5.10 \pm 0.23$ & $1.26 \pm 0.02$ & $9.10 \pm 0.73$ & $2.58 \pm 0.46$ \\
\hline
\end{tabular}
\end{table*}
\begin{figure*}
\centering
\includegraphics[width = 18cm]{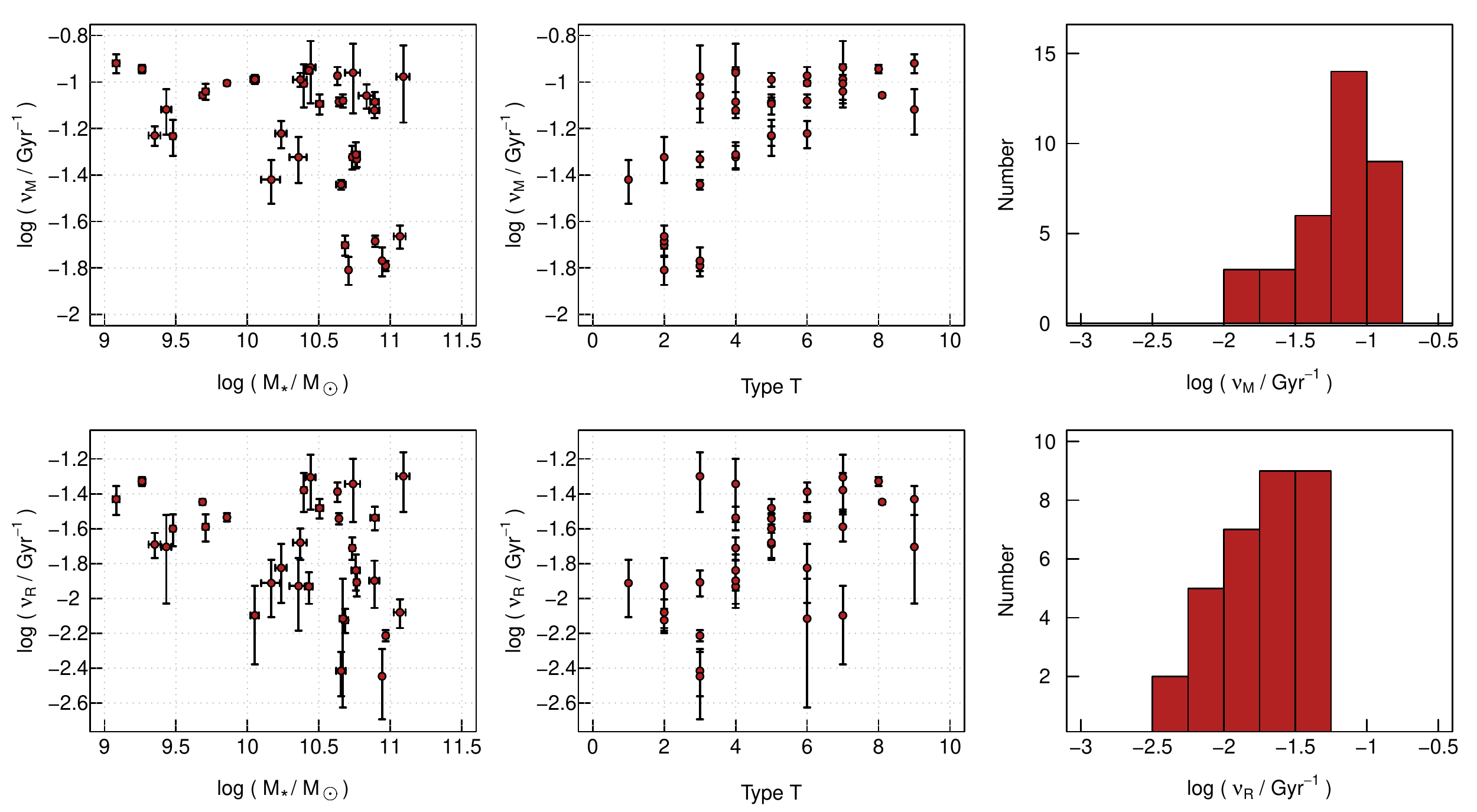}
\caption{The specific mass and radial growth rates $\nu_\textrm{M}$ (top) and $\nu_\textrm{R}$ (bottom) as a function of disc stellar mass (left) and morphological type (middle) and the relative histograms (right). Lower panels contain only the 32/35 galaxies with $\nu_\textrm{R} > 0$. Error bars are formal fitting uncertainties. The distributions of $\nu_\textrm{M}$ and $\nu_\textrm{R}$ have some similarities (see text), but $\nu_\textrm{R}$ values are sistematically lower by $\sim 0.5$ dex.}\label{fig::results}
\end{figure*}

\section{Results}\label{sec::results}
The results of our analysis are listed in Table \ref{table::results}. The quoted errors are just the formal fitting ones; in particular, they do not take into account systematic uncertainties, which, as discussed in Sec. \ref{sec::systematics}, might be important for the structural parameters $M_\star$ and $R_\star$, but have a limited impact on the growth parameters $\nu_\textrm{M}$ and $\nu_\textrm{R}$.

\subsection{Inside-out growth}\label{sec::ResultsInsideOut}
While our analysis is able to reveal both positive and negative radial growth rates, we find that 32 galaxies, out of 35, show $\nu_\textrm{R} > 0$. Of the remaining 3 galaxies with a formally negative radial growth rate, two (NGC 4569 and NGC 4826) have more than $100\%$ uncertainty in $\nu_\textrm{R}$ and hence are consistent with evolution of the stellar scalelength in one sense or the other, or with no evolution. Incidentally, we point out that both these galaxies are known to have peculiar properties: NGC 4569 is an anemic spiral in the Virgo cluster, probably significantly affected by ram pressure stripping (\citealt{Boselli+06}), while NGC 4826 is likely to have undergone a strongly misaligned merger, as suggested by the presence of a counter-rotating gaseous disc in the outskirts (\citealt{Braun+94}). For only one galaxy in our sample, NGC 1097, we clearly find the signature of a shrinking of the disc. It is interesting to notice, \emph{a posteriori}, that this galaxy has a very disturbed morphology. This is likely due to a strong interaction with the companion NGC 1097A \footnote{Note that NGC1097A cannot be seen in our online Atlas, since it has been masked out from our $3.6\; \mu \textrm{m}$ map.}. This object is listed as a peculiar elliptical in the RC3 catalogue (\citealt{3RC}),
though with `uncertain' classification.
In the GALEX Atlas (\citealt{GALEXAtlas}), NGC 1097A is clearly visible as a clump northwest of the prominent bar of NGC 1097, bright in NIR and NUV, faint in FUV and surrounded by an extended, FUV bright, disc-like structure, all properties common, in the GALEX Atlas, to the bulges of spiral galaxies. This is suggestive that the whole system may be a galaxy pair in an advanced stage of merging. We also noted that NGC 1097 is the object with the largest derived disc scalelength (6.32 kpc) and we verified that this result is not changed if we exclude from the exponential fit the whole radial range occupied by NGC 1097A. NGC 1097 has also been suggested to have undergone other significant interactions in the recent past (\citealt{HigdonWallin2003}). If our interpretation of a strong interaction state, likely a merger, for this system, is correct, then there is no surprise that it behaves differently from the regular evolution of isolated galaxies.

From these considerations, we can conclude that our findings are in excellent agreement with the general predictions of the inside-out growth scenario for the evolution of spiral galaxies. 

\subsection{Mass and radial growth rates}\label{sec::ResultsGrowthRates}
In Fig. \ref{fig::results} the mass and radial growth rates of the galaxies in our sample are plotted against disc stellar mass and morphological type. Since we are using logarithmic units, radial growth rates (lower panels) are reported only for those 32/35 galaxies with $\nu_\textrm{R} > 0$ (see Sec. \ref{sec::ResultsInsideOut}). Error bars represent formal fitting errors only. In particular, errors on distance and inclination are not taken into account in these plots. As discussed in Sec. \ref{sec::systematics}, these additional errors can affect stellar masses, but not $\nu_\textrm{M}$ and $\nu_\textrm{R}$, which are only subject to a common multiplicative uncertainty due to global calibration issues. 
We also give in Fig. \ref{fig::results} the histograms for the distributions of $\nu_\textrm{M}$ and $\nu_\textrm{R}$, binned in logarithmic intervals of 0.25 dex width.

From the upper panels of Fig. \ref{fig::results} we see that the specific mass growth rates (or specific star formation rates, or sSFR) of the discs of our galaxies take a relatively narrow range of values, with most of our points clustered around $\nu_\textrm{M} \sim 0.1\; \textrm{Gyr}^{-1}$, which corresponds to a mass growth timescale of $\sim 10 \; \textrm{Gyr}$.
This is in substantial agreement with the typical sSFR of star-forming galaxies in the Local Universe (e.g. \citealt{Elbaz+11}) and in particular with the relative constancy of the sSFR of the discs of spiral galaxies (\citealt{Abramson+14}), although a more detailed comparison would require more statistics and a careful treatment of global systematics (Sec. \ref{sec::systematics}), which is beyond the scope of this work (see e.g. \citealt{Speagle+14} about subtle issues concerning the homogeneization of measurements of these kind).
We find that a small group of 6 galaxies (NGC 2841, NGC 3031, NGC 4569, NGC 4579, NGC 4826, NGC 4725) have a particularly low sSFR, with $\log (\nu_\textrm{M}/\textrm{Gyr}^{-1}) < -1.5$. 
We have already recognized two of them (NGC 4569 and NGC 4826) as objects with peculiar properties and a close to vanishing radial growth rate (see Sec. \ref{sec::ResultsInsideOut}), but we cannot tell whether these peculiarities have a direct physical connection with the low measured values of $\nu_\textrm{M}$. However, we can see from the upper-mid panel that the whole group of 6 slowly-evolving galaxies are also among the galaxies of the earliest types in our sample. This may be interpreted as an indication of downsizing (e.g. \citealt{Cowie+96}): galaxies with high mass and early types are more likely to have completed most of their evolution in ancient epochs and hence to be growing with only mild rates nowadays. Also, galaxies of high mass and early-type might be more subject to star-formation quenching, the origin of which and its connection with morphology is still matter of investigation (e.g \citealt{Martig+09}; \citealt{Pan+14}).

In the lower panels of Fig. \ref{fig::results} we can see the distribution of the radial growth rates, which are the main novelty of this work.
When plotted against disc mass, the radial growth rate $\nu_\textrm{R}$ shows a quite similar distribution with respect to the one of $\nu_\textrm{M}$, but systematically shifted downwards by $\sim0.5 \; \textrm{dex}$. This suggests that our galaxies are growing in size, on average, at about 1/3 of the rate at which they are growing in stellar mass. The histogram of $\nu_\textrm{R}$ reveals a distribution that is similarly asymmetric, though less strongly peaked, with respect to the one of $\nu_\textrm{M}$. More than 50\% of our galaxies are in the two bins around $\log(\nu_\textrm{R}/\textrm{Gyr}^{-1}) = -1.5$, that is 0.5 dex below the peak of the $\nu_\textrm{M}$ distribution, corresponding to a typical radial growth timescale of $\sim 30 \; \textrm{Gyr}$.
Since the radial growth rate of galaxy discs has been studied much less than the sSFR, it is less obvious to compare our findings with the ones of previous studies.
However, we notice the typical timescale reported above is compatible with a radial growth of $\sim 25 \%$ in the last $\sim 7 \; \textrm{Gyr}$, very similar to what found by \cite{MMIII}. Our typical $\nu_\textrm{R}$ is also consistent with the results from the detailed study of resolved colour-magnitude diagrams for M 33 (\citealt{Williams+09}), while NGC 300 seems to have been growing at about half of this rate (\citealt{Gogarten+10}). However, our estimates strictly refer to the current time and caution is mandatory in extrapolating these instantaneous measurements to a significant fraction of the past history of galaxies.

\begin{table}
\centering
\caption{Basic statistics for our derived specific mass and radial growth rates (cfr. Fig. \ref{fig::results}). Note that discs with higher masses have lower median values and a higher scatter for both $\nu_\textrm{M}$ and $\nu_\textrm{R}$.}\label{table::statistics}
\begin{tabular}{lccc}
  \hline
    & $M_\star < 10^{10} M_\Sun$ & $M_\star > 10^{10} M_\Sun$ & All  \\
\hline
$\log(\nu_\textrm{M}/\textrm{Gyr}^{-1})$ & & &\\
Median & -1.05 & -1.12 & -1.09  \\
Scatter & \;0.13 & \;0.25 & \;0.20 \\
& & & \\
$\log(\nu_\textrm{R}/\textrm{Gyr}^{-1})$ & & & \\
Median &  -1.56 & -1.87 & -1.70 \\
Scatter & \;0.18 & \;0.37 & \;0.35 \\
\hline
\end{tabular}
\end{table}

Apart from the global vertical shift, the distributions of galaxies in the upper-left and bottom-left panels of Fig. \ref{fig::results} have a quite well defined common shape. In both cases, there is a horizontal upper envelope, close to the peak of the distribution, and a continuous increase of the scatter with increasing disc mass. Such a scatter is asymmetric and biased towards low values of $\nu_\textrm{M}$ and $\nu_\textrm{R}$, with the result that, on average, more massive discs appear to  grow at a slower rate, both in mass and size, than the less massive ones. This effect is quantified in Table \ref{table::statistics}, where the median mass and radial growth rates are reported, together with the associated scatter, for two subsamples with a derived disc mass lower or greater than $10^{10} \; \textrm{M}_\Sun$. 
It is not easy to understand, with our relatively small sample, if some simple physical property can be invoked to explain the scatter at high disc masses. However, we performed some simple checks and did not find any particular correlation between the position of galaxies in our plots and special properties, including the presence of a bar, an XUV disc, a break in the exponential profile, or indications of a warp or an interaction. We are therefore tempted to interpret the effect as intrinsic.

\section{Implications for the evolution of scaling relations of disc galaxies}\label{sec::scalingrelations}
{}
\subsection{The mass-radial growth connection}
Among our derived quantities, the mass and radial growth rates are the less affected by systematic uncertainties (see Sec. \ref{sec::systematics}). Hence, the most reliable of our results are those that we can derive plotting $\nu_\textrm{M}$ and $\nu_\textrm{R}$ against each other. This is also important to understand whether the shift, that we found in Sec. \ref{sec::ResultsGrowthRates}, of a factor $\sim 3$ between $\nu_\textrm{M}$ and $\nu_\textrm{R}$ is significant only at a statistical level or if it reflects an evolutionary property of individual galaxies.
\begin{figure}
\centering
\includegraphics[width=9cm]{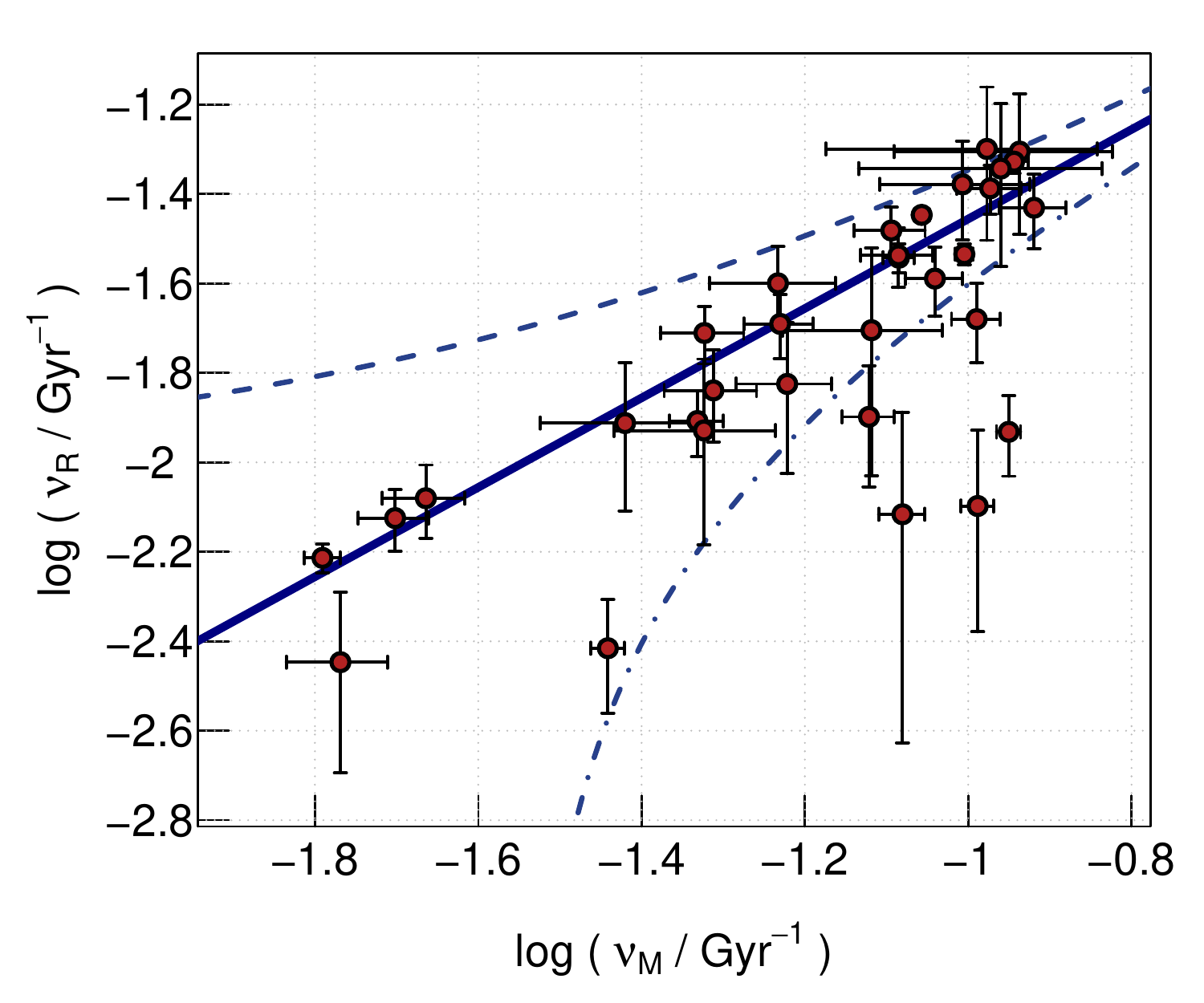}
\caption{The relation between the specific mass growth rate and the specific radial growth rate of galaxy discs. The points are the results of our measurements, the lines are predictions of some simple models. The solid line is the expectation if known scaling relations of disc galaxies are not evolving with time, the dahed line and the dot-dashed line are for scaling relations evolving on a timescale of $100 \; \textrm{Gyr}$, in one sense or the other (see text for details). A more rapid evolution is excluded by our results.
}\label{fig::GrowthPlot}
\end{figure}

Indeed, this experiment (Fig. \ref{fig::GrowthPlot}) reveals that the two growth rates are related to each other much more strongly than they are, individually, with mass or morphological type (cfr. Fig. \ref{fig::results}).

The fact that masses and sizes of galaxies grow in an interlinked way is not very surprising on its own. Hence, rather than fitting a straight line to the points in Fig. \ref{fig::GrowthPlot}, we prefer to seek for some simple physical explanation that can give a quantitative account to our finding.

\subsection{A comparison with a simple theoretical prediction}\label{sec::Comparison}
Let us assume that mass and size of the discs of spiral galaxies are connected by a power-law (e.g. \citealt{Courteau+07}, \citealt{Lange+15}):
\begin{equation}\label{powerlaw}
R_\star = AM_\star^{\alpha}
\end{equation}
Furthermore, let us assume that the coefficients $A$ and $\alpha$ are not evolving with time, so that the relation \eqref{powerlaw} defines not only the present locus, but also the evolutionary track of stellar discs. Then it immediately follows that the specific mass and radial growth rates should be linked by the very simple linear relation:
\begin{equation}\label{generalprediction}
\nu_\textrm{R} = \alpha \nu_\textrm{M}
\end{equation}
or, in logarithmic units:
\begin{equation}\label{generalpredictionlog}
\log \nu_\textrm{R} = \log \alpha + \log \nu_\textrm{M}
\end{equation}
Independently on the value of $\alpha$, equation \eqref{generalpredictionlog} implies that $(\nu_\textrm{M}, \nu_\textrm{R})$ points should lie, in a double logarithmic plot like Fig. \ref{fig::GrowthPlot}, on a line of unitary slope; this is indeed the slope of the solid line drawn in Fig. \ref{fig::GrowthPlot}, which gives a quite good account for the distribution of our datapoints. This is already suggestive that our results are consistent with the existence of a non-evolving, power-law, mass-size relation for the discs of spiral galaxies.

Of course, for our simple scenario to be fully predictive, not only the slope, but also the intercept, of such a straight line should be predicted as well, which is accomplished by specifying the expected value for $\alpha$.
To do this, we just combine two well-known scaling relations for disc galaxies, the Tully-Fisher relation (\citealt{McGaugh}), between the rotation velocity $V$ and the mass $M$ of a spiral galaxy:
\begin{equation}\label{TF}
V \; \propto \; M^{0.25}
\end{equation}
and the Fall relation (\citealt{RF12}), between specific angular momentum $l$ and mass:
\begin{equation}\label{Fall}
l \; \propto \; M^{0.6}
\end{equation}
Since exponential discs belong to a structurally self-similar family, one also has:
\begin{equation}\label{selfsimilar}
R_\star \; \propto \; \frac{l}{V}
\end{equation}
Substituting \eqref{Fall} and \eqref{TF} into \eqref{selfsimilar} we get a power-law mass-size relation of the form \eqref{powerlaw}, with $\alpha = 0.35$. 
This is not far from the value 0.32 empirically derived by \cite{Courteau+07} as an average slope for the mass-scalelength relation of disc galaxies in the Local Universe (since it was derived in the I-band, residual effects cannot be excluded arising from M/L variations). Shallower slopes are frequently found by studies based on half-light radius rather than disc scalelength (see e.g. \citealt{Lange+15}); this is in qualitative agreement with expectations if we consider an obvious morphological effect (more massive galaxies tend to have more prominent bulges and hence a smaller ratio between half-light radius and scalelength).
 
To define our simple model, we just retained the value $\alpha = 0.35$, derived from the Tully-Fisher and the Fall relations as explained above, and we adopted it to draw the solid line in Fig. \ref{fig::GrowthPlot}. Hence we see that the majority of our data-points lie on a locus that can be independently predicted, without any free parameter, just assuming that known scaling relations for disc galaxies hold and are not evolving with time. Since these simple hypotheses are completely independent from the way our results were derived, the agreement between the two is very unlikely to occur by chance, or to be due to biases of any kind, and we are tempted to interpret this finding as an indication for the validity of both our method and the hypotheses themselves.

We can also consider what effect residual systematics on $\nu_\textrm{M}$ and $\nu_\textrm{R}$ could have on our findings. As discussed in Sec. \ref{sec::systematics}, the effect is an unknown common multiplicative factor for $\nu_\textrm{M}$ and $\nu_\textrm{R}$. In the diagram shown in Fig. \ref{fig::GrowthPlot}, this implies a collective motion of all points along a line of unitary slope, or, equivalently, a mapping of the theoretical line into itself. Therefore, our conclusions are robust at least against the most obvious systematic uncertainties.

A word of caution is appropriate, however, against possible overinterpretation of our model and result. In fact, the Tully-Fisher relation is known to hold better for the whole baryonic content of spiral galaxies (\citealt{McGaugh}), while we have impicitly applied it just to the stellar mass of the disc. On the other side, the Fall relation seems to hold better when the disc component is considered separately from the bulge. Hence, it can be argued that, in deriving our predictions, we have mixed non-homogeneous empirical evidence. A more detailed analysis, taking this aspect into a proper account, would be interesting, but is beyond the scope of this work.

\subsection{Evolutionary effects}
In the previous Section we have seen that our results are compatible with the Tully-Fisher and Fall relations to be not evolving with time. To quantify this statement, we put here an upper limit on how fast a possible evolution can be in order to be still compatible with our results. For simplicity, we focus our attention on the evolution in normalization, although a similar analysis could be performed for the slope evolution as well.

If, in \eqref{powerlaw}, we allow the normalization $A$ to change with time, then \eqref{generalprediction} simply modifies into:
\begin{equation}\label{modifiedgeneralprediction}
\nu_\textrm{R} = \nu_\textrm{A} + \alpha\nu_\textrm{M}
\end{equation}
where:
\begin{equation}\label{nuAdef}
\nu_A (t) := \frac{d}{dt}( \ln A)(t) = \frac{\dot{A}(t)}{A(t)}
\end{equation}
is the specific growth rate of the normalization coefficient $A$. Of course, when equation \eqref{modifiedgeneralprediction} is compared with our observations in the Local Universe, $\nu_\textrm{A}$ has to be intended as evaluated at the present time.

The dashed and dot-dashed lines in Fig. \ref{fig::GrowthPlot} show the predictions of two models with a very mild evolution in normalization, in one sense or the other: $\nu_\textrm{A} = \pm \; 0.01 \; \textrm{Gyr}^{-1}$. It is clearly seen that the predicted distribution of galaxies in the $(\nu_\textrm{M}, \nu_\textrm{R})$ plane is extremely sensitive to the parameter $\nu_\textrm{A}$, making this diagram a new and powerful observational tool to constrain the evolution of scaling relations of galaxy discs. 
Also, since the two additional lines are both inconsistent with the empirical distribution, we quantitatively infer that, even admitting that scaling relations are evolving with time, they are doing so on timescales that are larger than $(\nu_\textrm{A, max})^{-1} = 100 \; \textrm{Gyr}$, hence much larger than the Hubble time.

Strictly speaking, the statement above mainly refers to the mass-size relation \eqref{powerlaw}. In fact, although the Tully-Fisher relation \eqref{TF} and the Fall relation \eqref{Fall} are the backbone of the simple model sketched in \ref{sec::Comparison}, it may be considered not trivial to draw out conclusions concerning them individually, since they both involve kinematics, while we did not directly make use of kinematical data. However, our results indicate that an evolution of the Tully-Fisher relation, if present, has to be accompanied and finely balanced by an opposite evolution of the Fall relation.

If compared with the direct observational study of scaling relations of disc galaxies at different redshifts (with all the appropriate \emph{caveats}, see Sec. \ref{sec::Introduction}), our results are in better agreement with those finding little or no evolution (see again references in Sec. \ref{sec::Introduction}), for either the mass-size or the Tully-Fisher relation, while, to our knowledge, no similar studies are available yet concerning the Fall relation. However, we stress again that our empirical upper limit ($|\nu_\textrm{A}| < 0.01 \; \textrm{Gyr}^{-1}$) only refers to $\nu_\textrm{A}$ evaluated at the present time and hence it is related (though not equal) to the slope of empirical $A(z)$ relations as measured at $z=0$. For instance, our results have no formal tension with those of \cite{Trujillo+06}, who still find a preference for evolutionary models, since their data points are also perfectly consistent with the slope of the $A(z)$ relation to vanish up to $z\sim1$;
to tell the difference between models, more precise measurements at moderate redshift, or the use of a sensitive local diagnostics like the one that we have proposed here, can be a valuable complement to pioneering observational campaigns in the extremely distant Universe.

\section{Summary}\label{sec::Summary}
In this work, we have developed, from very simple assumptions, a model that predicts a universal shape for the radial profile of the star formation rate surface density (SFRD) of spiral galaxies. This model accounts for the basic properties of observed profiles and naturally includes a parametrization of the growth of stellar discs. As a consequence, we have devised a novel, simple and powerful method to measure the instantaneous mass and radial growth rates of stellar discs, based on their SFRD profiles. We have applied our method to a sample of 35 nearby spiral galaxies. Our main results are:

\begin{enumerate}
\item For most of the galaxies in our sample, the SFRD profile is satisfactorily reproduced by our model, in such a way that we could measure the mass and radial growth rates $\nu_\textrm{M}$ and $\nu_\textrm{R}$ of their stellar discs.
\item Virtually all galaxies show the signature of inside-out growth ($\nu_\textrm{R} > 0$).
\item Typical timescales for the mass and radial growth of our stellar discs are of the order of $ \sim10 \; \textrm{Gyr}$ and $\sim 30 \; \textrm{Gyr}$, respectively, with some uncertainty due to systematic effects.
\item The mass and radial growth rates appear to obey a simple linear relation, with galaxy discs growing in size at $\sim 0.35$ times the rate at which they grow in mass. Compared with the individual timescales given above, this dimensionless ratio is more robust against systematic uncertainties.
\item The distribution of galaxies in the $(\nu_\textrm{M},\nu_\textrm{R})$ plane is a sensitive diagnostics for the evolution of scaling relations of galaxy discs.
\item Our results are in very good agreement with a simple model, without free parameters, based on the universality of the Tully-Fisher relation and the Fall relation, suggesting that they are not evolving with time. Possible residual evolution is constrained to occur on timescales that are much larger than the age of the Universe.  
\end{enumerate}

\section*{Acknowledgments}
G.P. is grateful to L. Posti and E. M. Di Teodoro for useful discussion. G.P. and F.F. acknowledge financial support from PRIN MIUR 2010-2011, project `The Chemical and Dynamical Evolution of the Milky Way and Local Group Galaxies', prot. 2010LY5N2T. In this work we made use of the Groningen Image Processing System.

\bibliographystyle{mn2e}
\bibliography{mybib}{}

\begin{thebibliography}{84}
\expandafter\ifx\csname natexlab\endcsname\relax\def\natexlab#1{#1}\fi

\bibitem[{{Abramson} {et~al}\mbox{.}(2014){Abramson}, {Kelson}, {Dressler},
  {Poggianti}, {Gladders}, {Oemler}, \& {Vulcani}}]{Abramson+14}
{Abramson} L.~E., {Kelson} D.~D., {Dressler} A., {Poggianti} B., {Gladders}
  M.~D., {Oemler}, Jr. A., {Vulcani} B., 2014, \apjl, 785, L36

\bibitem[{{Aumer} \& {Binney}(2009)}]{AB09}
{Aumer} M., {Binney} J.~J., 2009, \mnras, 397, 1286

\bibitem[{{Barden} {et~al}\mbox{.}(2005){Barden}, {Rix}, {Somerville}, {Bell},
  {H{\"a}u{\ss}ler}, {Peng}, {Borch}, {Beckwith}, {Caldwell}, {Heymans},
  {Jahnke}, {Jogee}, {McIntosh}, {Meisenheimer}, {S{\'a}nchez}, {Wisotzki}, \&
  {Wolf}}]{Barden+05}
{Barden} M. {et~al.}, 2005, \apj, 635, 959

\bibitem[{{Barker} {et~al}\mbox{.}(2011){Barker}, {Ferguson}, {Cole}, {Ibata},
  {Irwin}, {Lewis}, {Smecker-Hane}, \& {Tanvir}}]{Barker+11}
{Barker} M.~K., {Ferguson} A.~M.~N., {Cole} A.~A., {Ibata} R., {Irwin} M.,
  {Lewis} G.~F., {Smecker-Hane} T.~A., {Tanvir} N.~R., 2011, \mnras, 410, 504

\bibitem[{{Bell} \& {de Jong}(2000)}]{BelldeJong00}
{Bell} E.~F., {de Jong} R.~S., 2000, \mnras, 312, 497

\bibitem[{{Bell} {et~al}\mbox{.}(2003){Bell}, {McIntosh}, {Katz}, \&
  {Weinberg}}]{Bell+03}
{Bell} E.~F., {McIntosh} D.~H., {Katz} N., {Weinberg} M.~D., 2003, \apjs, 149,
  289

\bibitem[{{Boissier} \& {Prantzos}(1999)}]{BP99}
{Boissier} S., {Prantzos} N., 1999, \mnras, 307, 857

\bibitem[{{Boissier} {et~al}\mbox{.}(2007){Boissier}, {Gil de Paz}, {Boselli},
  {Madore}, {Buat}, {Cortese}, {Burgarella}, {Mu{\~n}oz-Mateos}, {Barlow},
  {Forster}, {Friedman}, {Martin}, {Morrissey}, {Neff}, {Schiminovich},
  {Seibert}, {Small}, {Wyder}, {Bianchi}, {Donas}, {Heckman}, {Lee},
  {Milliard}, {Rich}, {Szalay}, {Welsh}, \& {Yi}}]{Boissier+07}
{Boissier} S. {et~al.}, 2007, \apjs, 173, 524

\bibitem[{{Boissier} {et~al}\mbox{.}(2008){Boissier}, {Gil de Paz}, {Boselli},
  {Buat}, {Madore}, {Chemin}, {Balkowski}, {Amram}, {Carignan}, \& {van
  Driel}}]{Boissier+08}
{Boissier} S. {et~al.}, 2008, \apj, 681, 244

\bibitem[{{Boselli} {et~al}\mbox{.}(2006){Boselli}, {Boissier}, {Cortese}, {Gil
  de Paz}, {Seibert}, {Madore}, {Buat}, \& {Martin}}]{Boselli+06}
{Boselli} A., {Boissier} S., {Cortese} L., {Gil de Paz} A., {Seibert} M.,
  {Madore} B.~F., {Buat} V., {Martin} D.~C., 2006, \apj, 651, 811

\bibitem[{{Braun} {et~al}\mbox{.}(1994){Braun}, {Walterbos}, {Kennicutt}, \&
  {Tacconi}}]{Braun+94}
{Braun} R., {Walterbos} R.~A.~M., {Kennicutt}, Jr. R.~C., {Tacconi} L.~J.,
  1994, \apj, 420, 558

\bibitem[{{Briggs}(1990)}]{Briggs90}
{Briggs} F.~H., 1990, \apj, 352, 15

\bibitem[{{Buat} {et~al}\mbox{.}(2005){Buat}, {Iglesias-P{\'a}ramo}, {Seibert},
  {Burgarella}, {Charlot}, {Martin}, {Xu}, {Heckman}, {Boissier}, {Boselli},
  {Barlow}, {Bianchi}, {Byun}, {Donas}, {Forster}, {Friedman}, {Jelinski},
  {Lee}, {Madore}, {Malina}, {Milliard}, {Morissey}, {Neff}, {Rich},
  {Schiminovitch}, {Siegmund}, {Small}, {Szalay}, {Welsh}, \& {Wyder}}]{B05}
{Buat} V. {et~al.}, 2005, \apjl, 619, L51

\bibitem[{{Buitrago} {et~al}\mbox{.}(2008){Buitrago}, {Trujillo}, {Conselice},
  {Bouwens}, {Dickinson}, \& {Yan}}]{Buitrago+08}
{Buitrago} F., {Trujillo} I., {Conselice} C.~J., {Bouwens} R.~J., {Dickinson}
  M., {Yan} H., 2008, \apjl, 687, L61

\bibitem[{{Case} \& {Bhattacharya}(1998)}]{CB98}
{Case} G.~L., {Bhattacharya} D., 1998, \apj, 504, 761

\bibitem[{{Chiappini}, {Matteucci} \& {Romano}(2001){Chiappini}, {Matteucci},
  \& {Romano}}]{Chiappini}
{Chiappini} C., {Matteucci} F., {Romano} D., 2001, \apj, 554, 1044

\bibitem[{{Cortese} {et~al}\mbox{.}(2008){Cortese}, {Boselli}, {Franzetti},
  {Decarli}, {Gavazzi}, {Boissier}, \& {Buat}}]{C08}
{Cortese} L., {Boselli} A., {Franzetti} P., {Decarli} R., {Gavazzi} G.,
  {Boissier} S., {Buat} V., 2008, \mnras, 386, 1157

\bibitem[{{Courteau} {et~al}\mbox{.}(2007){Courteau}, {Dutton}, {van den
  Bosch}, {MacArthur}, {Dekel}, {McIntosh}, \& {Dale}}]{Courteau+07}
{Courteau} S., {Dutton} A.~A., {van den Bosch} F.~C., {MacArthur} L.~A.,
  {Dekel} A., {McIntosh} D.~H., {Dale} D.~A., 2007, \apj, 671, 203

\bibitem[{{Cowie} {et~al}\mbox{.}(1996){Cowie}, {Songaila}, {Hu}, \&
  {Cohen}}]{Cowie+96}
{Cowie} L.~L., {Songaila} A., {Hu} E.~M., {Cohen} J.~G., 1996, \aj, 112, 839

\bibitem[{{Dale} {et~al}\mbox{.}(2005){Dale}, {Bendo}, {Engelbracht}, {Gordon},
  {Regan}, {Armus}, {Cannon}, {Calzetti}, {Draine}, {Helou}, {Joseph},
  {Kennicutt}, {Li}, {Murphy}, {Roussel}, {Walter}, {Hanson}, {Hollenbach},
  {Jarrett}, {Kewley}, {Lamanna}, {Leitherer}, {Meyer}, {Rieke}, {Rieke},
  {Sheth}, {Smith}, \& {Thornley}}]{Dale+05}
{Dale} D.~A. {et~al.}, 2005, \apj, 633, 857

\bibitem[{{de Vaucouleurs} {et~al}\mbox{.}(1991){de Vaucouleurs}, {de
  Vaucouleurs}, {Corwin}, {Buta}, {Paturel}, \& {Fouqu{\'e}}}]{3RC}
{de Vaucouleurs} G., {de Vaucouleurs} A., {Corwin}, Jr. H.~G., {Buta} R.~J.,
  {Paturel} G., {Fouqu{\'e}} P., 1991, {Third Reference Catalogue of Bright
  Galaxies}. Springer-Verlag, New York

\bibitem[{{Elbaz} {et~al}\mbox{.}(2011){Elbaz}, {Dickinson}, {Hwang},
  {D{\'{\i}}az-Santos}, {Magdis}, {Magnelli}, {Le Borgne}, {Galliano},
  {Pannella}, {Chanial}, {Armus}, {Charmandaris}, {Daddi}, {Aussel}, {Popesso},
  {Kartaltepe}, {Altieri}, {Valtchanov}, {Coia}, {Dannerbauer}, {Dasyra},
  {Leiton}, {Mazzarella}, {Alexander}, {Buat}, {Burgarella}, {Chary}, {Gilli},
  {Ivison}, {Juneau}, {Le Floc'h}, {Lutz}, {Morrison}, {Mullaney}, {Murphy},
  {Pope}, {Scott}, {Brodwin}, {Calzetti}, {Cesarsky}, {Charlot}, {Dole},
  {Eisenhardt}, {Ferguson}, {F{\"o}rster Schreiber}, {Frayer}, {Giavalisco},
  {Huynh}, {Koekemoer}, {Papovich}, {Reddy}, {Surace}, {Teplitz}, {Yun}, \&
  {Wilson}}]{Elbaz+11}
{Elbaz} D. {et~al.}, 2011, \aap, 533, A119

\bibitem[{{Elmegreen} {et~al}\mbox{.}(2005){Elmegreen}, {Elmegreen},
  {Vollbach}, {Foster}, \& {Ferguson}}]{Elmegreen+05}
{Elmegreen} B.~G., {Elmegreen} D.~M., {Vollbach} D.~R., {Foster} E.~R.,
  {Ferguson} T.~E., 2005, \apj, 634, 101

\bibitem[{{Erwin}, {Pohlen} \& {Beckman}(2008){Erwin}, {Pohlen}, \&
  {Beckman}}]{Erwin+08}
{Erwin} P., {Pohlen} M., {Beckman} J.~E., 2008, \aj, 135, 20

\bibitem[{{Fall}(1983)}]{Fall}
{Fall} S.~M., 1983, in IAU Symposium, Vol. 100, Internal Kinematics and
  Dynamics of Galaxies, {Athanassoula} E., ed., pp. 391--398

\bibitem[{{Fathi} {et~al}\mbox{.}(2012){Fathi}, {Gatchell}, {Hatziminaoglou},
  \& {Epinat}}]{Fathi+12}
{Fathi} K., {Gatchell} M., {Hatziminaoglou} E., {Epinat} B., 2012, \mnras, 423,
  L112

\bibitem[{{Fraternali} \& {Tomassetti}(2012)}]{FT12}
{Fraternali} F., {Tomassetti} M., 2012, \mnras, 426, 2166

\bibitem[{{Freeman}(1970)}]{Freeman}
{Freeman} K.~C., 1970, \apj, 160, 811

\bibitem[{{Gil de Paz} {et~al}\mbox{.}(2007){Gil de Paz}, {Boissier}, {Madore},
  {Seibert}, {Joe}, {Boselli}, {Wyder}, {Thilker}, {Bianchi}, {Rey}, {Rich},
  {Barlow}, {Conrow}, {Forster}, {Friedman}, {Martin}, {Morrissey}, {Neff},
  {Schiminovich}, {Small}, {Donas}, {Heckman}, {Lee}, {Milliard}, {Szalay}, \&
  {Yi}}]{GALEXAtlas}
{Gil de Paz} A. {et~al.}, 2007, \apjs, 173, 185

\bibitem[{{Goddard}, {Kennicutt} \& {Ryan-Weber}(2010){Goddard}, {Kennicutt},
  \& {Ryan-Weber}}]{Goddard+10}
{Goddard} Q.~E., {Kennicutt} R.~C., {Ryan-Weber} E.~V., 2010, \mnras, 405, 2791

\bibitem[{{Gogarten} {et~al}\mbox{.}(2010){Gogarten}, {Dalcanton}, {Williams},
  {Ro{\v s}kar}, {Holtzman}, {Seth}, {Dolphin}, {Weisz}, {Cole}, {Debattista},
  {Gilbert}, {Olsen}, {Skillman}, {de Jong}, {Karachentsev}, \&
  {Quinn}}]{Gogarten+10}
{Gogarten} S.~M. {et~al.}, 2010, \apj, 712, 858

\bibitem[{{Gonz{\'a}lez Delgado} {et~al}\mbox{.}(2014){Gonz{\'a}lez Delgado},
  {P{\'e}rez}, {Cid Fernandes}, {Garc{\'{\i}}a-Benito}, {de Amorim},
  {S{\'a}nchez}, {Husemann}, {Cortijo-Ferrero}, {L{\'o}pez Fern{\'a}ndez},
  {S{\'a}nchez-Bl{\'a}zquez}, {Bekeraite}, {Walcher}, {Falc{\'o}n-Barroso},
  {Gallazzi}, {van de Ven}, {Alves}, {Bland-Hawthorn}, {Kennicutt}, {Kupko},
  {Lyubenova}, {Mast}, {Moll{\'a}}, {Marino}, {Quirrenbach}, {V{\'{\i}}lchez},
  \& {Wisotzki}}]{CALIFA}
{Gonz{\'a}lez Delgado} R.~M. {et~al.}, 2014, \aap, 562, A47

\bibitem[{{Higdon} \& {Wallin}(2003)}]{HigdonWallin2003}
{Higdon} J.~L., {Wallin} J.~F., 2003, \apj, 585, 281

\bibitem[{{Ichikawa}, {Kajisawa} \& {Akhlaghi}(2012){Ichikawa}, {Kajisawa}, \&
  {Akhlaghi}}]{Ichikawa+12}
{Ichikawa} T., {Kajisawa} M., {Akhlaghi} M., 2012, \mnras, 422, 1014

\bibitem[{{Kennicutt} \& {Evans}(2012)}]{KE12}
{Kennicutt} R.~C., {Evans} N.~J., 2012, \araa, 50, 531

\bibitem[{{Kennicutt} {et~al}\mbox{.}(2003){Kennicutt}, {Armus}, {Bendo},
  {Calzetti}, {Dale}, {Draine}, {Engelbracht}, {Gordon}, {Grauer}, {Helou},
  {Hollenbach}, {Jarrett}, {Kewley}, {Leitherer}, {Li}, {Malhotra}, {Regan},
  {Rieke}, {Rieke}, {Roussel}, {Smith}, {Thornley}, \& {Walter}}]{SINGS}
{Kennicutt}, Jr. R.~C. {et~al.}, 2003, \pasp, 115, 928

\bibitem[{{Kroupa}(2001)}]{Kroupa2001}
{Kroupa} P., 2001, \mnras, 322, 231

\bibitem[{{Kubryk}, {Prantzos} \& {Athanassoula}(2013){Kubryk}, {Prantzos}, \&
  {Athanassoula}}]{KPA13}
{Kubryk} M., {Prantzos} N., {Athanassoula} E., 2013, \mnras, 436, 1479

\bibitem[{{Lange} {et~al}\mbox{.}(2015){Lange}, {Driver}, {Robotham}, {Kelvin},
  {Graham}, {Alpaslan}, {Andrews}, {Baldry}, {Bamford}, {Bland-Hawthorn},
  {Brough}, {Cluver}, {Conselice}, {Davies}, {Haeussler}, {Konstantopoulos},
  {Loveday}, {Moffett}, {Norberg}, {Phillipps}, {Taylor},
  {L{\'o}pez-S{\'a}nchez}, \& {Wilkins}}]{Lange+15}
{Lange} R. {et~al.}, 2015, \mnras, 447, 2603

\bibitem[{{Larson}(1976)}]{Larson76}
{Larson} R.~B., 1976, \mnras, 176, 31

\bibitem[{{Leitner} \& {Kravtsov}(2011)}]{LK11}
{Leitner} S.~N., {Kravtsov} A.~V., 2011, \apj, 734, 48

\bibitem[{{Leroy} {et~al}\mbox{.}(2008){Leroy}, {Walter}, {Brinks}, {Bigiel},
  {de Blok}, {Madore}, \& {Thornley}}]{Leroy+08}
{Leroy} A.~K., {Walter} F., {Brinks} E., {Bigiel} F., {de Blok} W.~J.~G.,
  {Madore} B., {Thornley} M.~D., 2008, \aj, 136, 2782

\bibitem[{{Lilly} {et~al}\mbox{.}(2013){Lilly}, {Carollo}, {Pipino}, {Renzini},
  \& {Peng}}]{Lilly}
{Lilly} S.~J., {Carollo} C.~M., {Pipino} A., {Renzini} A., {Peng} Y., 2013,
  \apj, 772, 119

\bibitem[{{Ludwig} {et~al}\mbox{.}(2012){Ludwig}, {Pasquali}, {Grebel}, \&
  {Gallagher}}]{Ludwig+12}
{Ludwig} J., {Pasquali} A., {Grebel} E.~K., {Gallagher}, III J.~S., 2012, \aj,
  144, 190

\bibitem[{{MacArthur} {et~al}\mbox{.}(2004){MacArthur}, {Courteau}, {Bell}, \&
  {Holtzman}}]{MacArthur+04}
{MacArthur} L.~A., {Courteau} S., {Bell} E., {Holtzman} J.~A., 2004, \apjs,
  152, 175

\bibitem[{{Mao}, {Mo} \& {White}(1998){Mao}, {Mo}, \& {White}}]{MaoMoWhite98}
{Mao} S., {Mo} H.~J., {White} S.~D.~M., 1998, \mnras, 297, L71

\bibitem[{{Martig} {et~al}\mbox{.}(2009){Martig}, {Bournaud}, {Teyssier}, \&
  {Dekel}}]{Martig+09}
{Martig} M., {Bournaud} F., {Teyssier} R., {Dekel} A., 2009, \apj, 707, 250

\bibitem[{{Martin} {et~al}\mbox{.}(2005){Martin}, {Fanson}, {Schiminovich},
  {Morrissey}, {Friedman}, {Barlow}, {Conrow}, {Grange}, {Jelinsky},
  {Milliard}, {Siegmund}, {Bianchi}, {Byun}, {Donas}, {Forster}, {Heckman},
  {Lee}, {Madore}, {Malina}, {Neff}, {Rich}, {Small}, {Surber}, {Szalay},
  {Welsh}, \& {Wyder}}]{GALEX}
{Martin} D.~C. {et~al.}, 2005, \apjl, 619, L1

\bibitem[{{McGaugh}(2012)}]{McGaugh}
{McGaugh} S.~S., 2012, \aj, 143, 40

\bibitem[{{Meidt} {et~al}\mbox{.}(2012){Meidt}, {Schinnerer}, {Knapen},
  {Bosma}, {Athanassoula}, {Sheth}, {Buta}, {Zaritsky}, {Laurikainen},
  {Elmegreen}, {Elmegreen}, {Gadotti}, {Salo}, {Regan}, {Ho}, {Madore}, {Hinz},
  {Skibba}, {Gil de Paz}, {Mu{\~n}oz-Mateos}, {Men{\'e}ndez-Delmestre},
  {Seibert}, {Kim}, {Mizusawa}, {Laine}, \& {Comer{\'o}n}}]{Meidt+12}
{Meidt} S.~E. {et~al.}, 2012, \apj, 744, 17

\bibitem[{{Meidt} {et~al}\mbox{.}(2014){Meidt}, {Schinnerer}, {van de Ven},
  {Zaritsky}, {Peletier}, {Knapen}, {Sheth}, {Regan}, {Querejeta},
  {Mu{\~n}oz-Mateos}, {Kim}, {Hinz}, {Gil de Paz}, {Athanassoula}, {Bosma},
  {Buta}, {Cisternas}, {Ho}, {Holwerda}, {Skibba}, {Laurikainen}, {Salo},
  {Gadotti}, {Laine}, {Erroz-Ferrer}, {Comer{\'o}n}, {Men{\'e}ndez-Delmestre},
  {Seibert}, \& {Mizusawa}}]{Meidt+14}
{Meidt} S.~E. {et~al.}, 2014, \apj, 788, 144

\bibitem[{{Miller} {et~al}\mbox{.}(2011){Miller}, {Bundy}, {Sullivan}, {Ellis},
  \& {Treu}}]{Miller+11}
{Miller} S.~H., {Bundy} K., {Sullivan} M., {Ellis} R.~S., {Treu} T., 2011,
  \apj, 741, 115

\bibitem[{{Miller} {et~al}\mbox{.}(2012){Miller}, {Ellis}, {Sullivan}, {Bundy},
  {Newman}, \& {Treu}}]{Miller+12}
{Miller} S.~H., {Ellis} R.~S., {Sullivan} M., {Bundy} K., {Newman} A.~B.,
  {Treu} T., 2012, \apj, 753, 74

\bibitem[{{Minchev}, {Chiappini} \& {Martig}(2014){Minchev}, {Chiappini}, \&
  {Martig}}]{MCM14}
{Minchev} I., {Chiappini} C., {Martig} M., 2014, \aap, 572, A92

\bibitem[{{Moll{\'a}} \& {D{\'{\i}}az}(2005)}]{MD05}
{Moll{\'a}} M., {D{\'{\i}}az} A.~I., 2005, \mnras, 358, 521

\bibitem[{{Mu{\~n}oz-Mateos} {et~al}\mbox{.}(2007){Mu{\~n}oz-Mateos}, {Gil de
  Paz}, {Boissier}, {Zamorano}, {Jarrett}, {Gallego}, \& {Madore}}]{MM+07}
{Mu{\~n}oz-Mateos} J.~C., {Gil de Paz} A., {Boissier} S., {Zamorano} J.,
  {Jarrett} T., {Gallego} J., {Madore} B.~F., 2007, \apj, 658, 1006

\bibitem[{{Mu{\~n}oz-Mateos}
  {et~al}\mbox{.}(2009{\natexlab{a}}){Mu{\~n}oz-Mateos}, {Gil de Paz},
  {Boissier}, {Zamorano}, {Dale}, {P{\'e}rez-Gonz{\'a}lez}, {Gallego},
  {Madore}, {Bendo}, {Thornley}, {Draine}, {Boselli}, {Buat}, {Calzetti},
  {Moustakas}, \& {Kennicutt}}]{MMII}
{Mu{\~n}oz-Mateos} J.~C. {et~al.}, 2009{\natexlab{a}}, \apj, 701, 1965

\bibitem[{{Mu{\~n}oz-Mateos}
  {et~al}\mbox{.}(2009{\natexlab{b}}){Mu{\~n}oz-Mateos}, {Gil de Paz},
  {Zamorano}, {Boissier}, {Dale}, {P{\'e}rez-Gonz{\'a}lez}, {Gallego},
  {Madore}, {Bendo}, {Boselli}, {Buat}, {Calzetti}, {Moustakas}, \&
  {Kennicutt}}]{MMI}
{Mu{\~n}oz-Mateos} J.~C. {et~al.}, 2009{\natexlab{b}}, \apj, 703, 1569

\bibitem[{{Mu{\~n}oz-Mateos} {et~al}\mbox{.}(2011){Mu{\~n}oz-Mateos},
  {Boissier}, {Gil de Paz}, {Zamorano}, {Kennicutt}, {Moustakas}, {Prantzos},
  \& {Gallego}}]{MMIII}
{Mu{\~n}oz-Mateos} J.~C., {Boissier} S., {Gil de Paz} A., {Zamorano} J.,
  {Kennicutt}, Jr. R.~C., {Moustakas} J., {Prantzos} N., {Gallego} J., 2011,
  \apj, 731, 10

\bibitem[{{Naab} \& {Ostriker}(2006)}]{NO06}
{Naab} T., {Ostriker} J.~P., 2006, \mnras, 366, 899

\bibitem[{{Pan} {et~al}\mbox{.}(2014){Pan}, {Li}, {Lin}, {Wang}, \&
  {Kong}}]{Pan+14}
{Pan} Z., {Li} J., {Lin} W., {Wang} J., {Kong} X., 2014, \apjl, 792, L4

\bibitem[{{Peebles}(1969)}]{Peebles}
{Peebles} P.~J.~E., 1969, \apj, 155, 393

\bibitem[{{Pohlen} \& {Trujillo}(2006)}]{PohlenTrujillo06}
{Pohlen} M., {Trujillo} I., 2006, \aap, 454, 759

\bibitem[{{Ravindranath} {et~al}\mbox{.}(2004){Ravindranath}, {Ferguson},
  {Conselice}, {Giavalisco}, {Dickinson}, {Chatzichristou}, {de Mello}, {Fall},
  {Gardner}, {Grogin}, {Hornschemeier}, {Jogee}, {Koekemoer}, {Kretchmer},
  {Livio}, {Mobasher}, \& {Somerville}}]{Ravindranath+04}
{Ravindranath} S. {et~al.}, 2004, \apjl, 604, L9

\bibitem[{{Regan} {et~al}\mbox{.}(2004){Regan}, {Thornley}, {Bendo}, {Draine},
  {Li}, {Dale}, {Engelbracht}, {Kennicutt}, {Armus}, {Calzetti}, {Gordon},
  {Helou}, {Hollenbach}, {Jarrett}, {Kewley}, {Leitherer}, {Malhotra}, {Meyer},
  {Misselt}, {Morrison}, {Murphy}, {Muzerolle}, {Rieke}, {Rieke}, {Roussel},
  {Smith}, \& {Walter}}]{Regan+04}
{Regan} M.~W. {et~al.}, 2004, \apjs, 154, 204

\bibitem[{{Romanowsky} \& {Fall}(2012)}]{RF12}
{Romanowsky} A.~J., {Fall} S.~M., 2012, \apjs, 203, 17

\bibitem[{{Ro{\v s}kar} {et~al}\mbox{.}(2012){Ro{\v s}kar}, {Debattista},
  {Quinn}, \& {Wadsley}}]{Roskar+12}
{Ro{\v s}kar} R., {Debattista} V.~P., {Quinn} T.~R., {Wadsley} J., 2012,
  \mnras, 426, 2089

\bibitem[{{Sch{\"o}nrich} \& {Binney}(2009)}]{SB09}
{Sch{\"o}nrich} R., {Binney} J., 2009, \mnras, 396, 203

\bibitem[{{Sellwood}(2014)}]{SellwoodReview}
{Sellwood} J.~A., 2014, Reviews of Modern Physics, 86, 1

\bibitem[{{Sellwood} \& {Binney}(2002)}]{SB02}
{Sellwood} J.~A., {Binney} J.~J., 2002, \mnras, 336, 785

\bibitem[{{Shen} {et~al}\mbox{.}(2003){Shen}, {Mo}, {White}, {Blanton},
  {Kauffmann}, {Voges}, {Brinkmann}, \& {Csabai}}]{Shen+03}
{Shen} S., {Mo} H.~J., {White} S.~D.~M., {Blanton} M.~R., {Kauffmann} G.,
  {Voges} W., {Brinkmann} J., {Csabai} I., 2003, \mnras, 343, 978

\bibitem[{{Simard} {et~al}\mbox{.}(1999){Simard}, {Koo}, {Faber}, {Sarajedini},
  {Vogt}, {Phillips}, {Gebhardt}, {Illingworth}, \& {Wu}}]{Simard+99}
{Simard} L. {et~al.}, 1999, \apj, 519, 563

\bibitem[{{Speagle} {et~al}\mbox{.}(2014){Speagle}, {Steinhardt}, {Capak}, \&
  {Silverman}}]{Speagle+14}
{Speagle} J.~S., {Steinhardt} C.~L., {Capak} P.~L., {Silverman} J.~D., 2014,
  \apjs, 214, 15

\bibitem[{{Thilker} {et~al}\mbox{.}(2007{\natexlab{a}}){Thilker}, {Bianchi},
  {Meurer}, {Gil de Paz}, {Boissier}, {Madore}, {Boselli}, {Ferguson},
  {Mu{\~n}oz-Mateos}, {Madsen}, {Hameed}, {Overzier}, {Forster}, {Friedman},
  {Martin}, {Morrissey}, {Neff}, {Schiminovich}, {Seibert}, {Small}, {Wyder},
  {Donas}, {Heckman}, {Lee}, {Milliard}, {Rich}, {Szalay}, {Welsh}, \&
  {Yi}}]{Thilker+07a}
{Thilker} D.~A. {et~al.}, 2007{\natexlab{a}}, \apjs, 173, 538

\bibitem[{{Thilker} {et~al}\mbox{.}(2007{\natexlab{b}}){Thilker}, {Boissier},
  {Bianchi}, {Calzetti}, {Boselli}, {Dale}, {Seibert}, {Braun}, {Burgarella},
  {Gil de Paz}, {Helou}, {Walter}, {Kennicutt}, {Madore}, {Martin}, {Barlow},
  {Forster}, {Friedman}, {Morrissey}, {Neff}, {Schiminovich}, {Small}, {Wyder},
  {Donas}, {Heckman}, {Lee}, {Milliard}, {Rich}, {Szalay}, {Welsh}, \&
  {Yi}}]{Thilker+07b}
{Thilker} D.~A. {et~al.}, 2007{\natexlab{b}}, \apjs, 173, 572

\bibitem[{{Tinsley}(1980)}]{Tinsley80}
{Tinsley} B.~M., 1980, \fcp, 5, 287

\bibitem[{{Tojeiro} {et~al}\mbox{.}(2013){Tojeiro}, {Masters}, {Richards},
  {Percival}, {Bamford}, {Maraston}, {Nichol}, {Skibba}, \&
  {Thomas}}]{Tojeiro+13}
{Tojeiro} R. {et~al.}, 2013, \mnras, 432, 359

\bibitem[{{Trujillo} {et~al}\mbox{.}(2006){Trujillo}, {F{\"o}rster Schreiber},
  {Rudnick}, {Barden}, {Franx}, {Rix}, {Caldwell}, {McIntosh}, {Toft},
  {H{\"a}ussler}, {Zirm}, {van Dokkum}, {Labb{\'e}}, {Moorwood},
  {R{\"o}ttgering}, {van der Wel}, {van der Werf}, \& {van
  Starkenburg}}]{Trujillo+06}
{Trujillo} I. {et~al.}, 2006, \apj, 650, 18

\bibitem[{{Tully} \& {Fisher}(1977)}]{TF}
{Tully} R.~B., {Fisher} J.~R., 1977, \aap, 54, 661

\bibitem[{{Vogt} {et~al}\mbox{.}(1997){Vogt}, {Phillips}, {Faber}, {Gallego},
  {Gronwall}, {Guzm{\'a}n}, {Illingworth}, {Koo}, \& {Lowenthal}}]{Vogt+97}
{Vogt} N.~P. {et~al.}, 1997, \apjl, 479, L121

\bibitem[{{Wang} {et~al}\mbox{.}(2011){Wang}, {Kauffmann}, {Overzier},
  {Catinella}, {Schiminovich}, {Heckman}, {Moran}, {Haynes}, {Giovanelli}, \&
  {Kong}}]{Wang+11}
{Wang} J. {et~al.}, 2011, \mnras, 412, 1081

\bibitem[{{Williams} {et~al}\mbox{.}(2009){Williams}, {Dalcanton}, {Dolphin},
  {Holtzman}, \& {Sarajedini}}]{Williams+09}
{Williams} B.~F., {Dalcanton} J.~J., {Dolphin} A.~E., {Holtzman} J.,
  {Sarajedini} A., 2009, \apjl, 695, L15

\bibitem[{{Yoachim}, {Ro{\v s}kar} \& {Debattista}(2012){Yoachim}, {Ro{\v
  s}kar}, \& {Debattista}}]{Yoachim+12}
{Yoachim} P., {Ro{\v s}kar} R., {Debattista} V.~P., 2012, \apj, 752, 97

\bibitem[{{Zaritsky} {et~al}\mbox{.}(2014){Zaritsky}, {Courtois},
  {Mu{\~n}oz-Mateos}, {Sorce}, {Erroz-Ferrer}, {Comer{\'o}n}, {Gadotti}, {Gil
  de Paz}, {Hinz}, {Laurikainen}, {Kim}, {Laine}, {Men{\'e}ndez-Delmestre},
  {Mizusawa}, {Regan}, {Salo}, {Seibert}, {Sheth}, {Athanassoula}, {Bosma},
  {Cisternas}, {Ho}, \& {Holwerda}}]{Zaritsky+14}
{Zaritsky} D. {et~al.}, 2014, \aj, 147, 134

\end{thebibliography}
\end{document}